\documentclass[twocolumn]{aastex631}

\submitjournal{AJ}

\shorttitle{High-resolution Emission Spectroscopy of KELT-9b}
\shortauthors{A. Ridden-Harper, E. de Mooij, R. Jayawardhana et al.}
\graphicspath{{./}}

\begin{document}

\title{High-resolution Emission Spectroscopy of the Ultrahot Jupiter KELT-9b:\\ Little Variation in Day- and Nightside Emission Line Contrasts}

\correspondingauthor{Andrew Ridden-Harper}
\email{ariddenharper@lco.global}

\author[0000-0002-5425-2655]{Andrew Ridden-Harper}
\affiliation{Department of Astronomy and Carl Sagan Institute, Cornell University, Ithaca, NY 14853, USA}
\affiliation{Las Cumbres Observatory, 6740 Cortona Drive, Suite 102, Goleta, CA 93117, USA}

\author[0000-0001-6391-9266]{Ernst de Mooij}
\affiliation{Astrophysics Research Centre, School of Mathematics and Physics, Queen’s University Belfast, University Road,
Belfast BT7 1NN, United Kingdom}

\author[0000-0001-5349-6853]{Ray Jayawardhana}
\affiliation{Department of Astronomy, Cornell University, Ithaca, NY 14853, USA}

\author[0000-0002-9308-2353]{Neale Gibson}
\affiliation{School of Physics, Trinity College Dublin, Dublin 2, Ireland}

\author[0000-0002-2656-909X]{Raine Karjalainen}
\affiliation{Astronomical Institute, Czech Academy of Sciences, Fri$\check{c}$ova~298, 25165, Ond$\check{r}$ejov, Czech Republic}
\affiliation{Instituto de Astrofísica de Canarias, c/ Vía Láctea s/n E-38205 La Laguna, Tenerife, Spain}
\affiliation{Isaac Newton Group of Telescopes, Apartado de Correos 321, Santa Cruz de La Palma, E-38700, Spain}

\author[0000-0003-0751-3231]{Marie Karjalainen}
\affiliation{Astronomical Institute, Czech Academy of Sciences, Fri$\check{c}$ova~298, 25165, Ond$\check{r}$ejov, Czech Republic}

\newcommand\NumPermutations{1000~}

%% Note that the \and command from previous versions of AASTeX is now
%% depreciated in this version as it is no longer necessary. AASTeX 
%% automatically takes care of all commas and "and"s between authors names.

%% AASTeX 6.31 has the new \collaboration and \nocollaboration commands to
%% provide the collaboration status of a group of authors. These commands 
%% can be used either before or after the list of corresponding authors. The
%% argument for \collaboration is the collaboration identifier. Authors are
%% encouraged to surround collaboration identifiers with ()s. The 
%% \nocollaboration command takes no argument and exists to indicate that
%% the nearby authors are not part of surrounding collaborations.

%% Mark off the abstract in the ``abstract'' environment. 
\begin{abstract}
The transmission spectrum of the ultrahot Jupiter KELT-9b ($T_{eq}$ $\sim$ 4000 K) exhibits absorption by several metal species. We searched for atomic and molecular lines in its emission spectrum by observing partial phase curves with the CARMENES spectrograph ($R$ $\sim$ 80,000 $-$ 95,000). We find evidence for emission by Si I in the atmosphere of KELT-9b for the first time.  Additionally we find evidence for emission by Mg I and Ca II, which were previously detected in transmission, and confirmed earlier detections of Fe I emission. Conversely, we find no evidence for dayside emission from Al I, Ca I, Cr I, FeH, Fe II, K I, Li I, Mg II, Na I, OH, Ti I, TiO, V I, V II, VO, and Y I.  By employing likelihood mapping, we find indications of there being little variation in emission line contrast between the day- and nightsides --suggesting that KELT-9b may harbor iron emission on its nightside.  Our results demonstrate that high-resolution ground-based emission spectroscopy can provide valuable insights into exoplanet atmospheres.

\end{abstract}

%% Keywords should appear after the \end{abstract} command. 
%% The AAS Journals now uses Unified Astronomy Thesaurus concepts:
%% https://astrothesaurus.org
%% You will be asked to selected these concepts during the submission process
%% but this old "keyword" functionality is maintained in case authors want
%% to include these concepts in their preprints.

%% From the front matter, we move on to the body of the paper.
%% Sections are demarcated by \section and \subsection, respectively.
%% Observe the use of the LaTeX \label
%% command after the \subsection to give a symbolic KEY to the
%% subsection for cross-referencing in a \ref command.
%% You can use LaTeX's \ref and \label commands to keep track of
%% cross-references to sections, equations, tables, and figures.
%% That way, if you change the order of any elements, LaTeX will
%% automatically renumber them.
%%
%% We recommend that authors also use the natbib \citep
%% and \citet commands to identify citations.  The citations are
%% tied to the reference list via symbolic KEYs. The KEY corresponds
%% to the KEY in the \bibitem in the reference list below. 

\section{Introduction} \label{sec:intro}

%%% https://journals.aas.org/revision-history/#_6.3.1
%% \textbf{text}
%% \deleted{text}
%% \replaced{old text}{new text}	

Ultrahot Jupiters have short orbital periods and hot host stars, leading to equilibrium temperatures $\gtrsim$2000 K.  The high temperatures result in their atmospheres being well described by equilibrium chemistry, and known condensates being confined to the nightside of the planets \citep{Kitzmann2018,Lothringer2018,Parmentier2018, Helling2019}.  

KELT-9b is the most extreme ultrahot Jupiter known, with an equilibrium temperature of about 4000 K \citep{Gaudi2017}, a dayside brightness temperature of 4600 K, and a nightside temperature of 3000 K \citep{Mansfield2020,Wong2020,Wong2021,Jones2022}.  Its extreme temperature has motivated several observational studies with ground- and space-based telescopes.  Ground-based high-resolution spectroscopy can identify atmospheric constituents unambiguously by resolving their spectral lines \citep[e.g.,][]{Snellen2010}.  Such observations have revealed the presence of several atomic species in KELT-9b's transmission spectrum, which probes low pressures at the day-night terminator. Specifically, these species are neutral hydrogen \citep{Yan2018,Cauley2019,Wyttenbach2020} and the metals Ca I, Ca II, Co, Cr I, Cr II, Fe I, Fe II, Mg I, Na I, Sc II, Sr II, Ti II, and Y II \citep[e.g.,][]{Hoeijmakers2018a,Hoeijmakers2019,Yan2019,Turner2020,Bello-Arufe2022,PaiAsnodkar2022}.  \cite{Pino2020} have also identified neutral iron in the dayside {\it emission} spectrum of KELT-9b.  Their observations determined that the atmosphere has a temperature inversion, and that the iron emission originates at the 10$^{-3}$ $-$10$^{-5}$ bar level.

KELT-9b's dayside emission spectrum has also been observed by the Hubble Space Telescope/Wide Field Camera 3 \citep{Changeat2021}.  Atmospheric retrievals with the TauREx3 code revealed evidence for emission by TiO, VO, FeH, and H$^-$. This is somewhat surprising because all molecules are thought to have thermally dissociated by around 4000 K \citep{Kitzmann2018}, which matches KELT-9b's dayside temperature.  The authors suggest that disequilibrium processes such as horizontal or vertical mixing may play a role in distributing these molecules to the dayside.

\citet{Kasper2021} observed the emission spectrum of KELT-9b with MAROON-X on Gemini-North and investigated its atmospheric composition with a high-resolution spectral retrieval, similar to that of \citet{Brogi2019}. Their retrieval considered H$^-$, Fe I, Fe II, Ti I, Ti II, Ca II, Mg I, and TiO. While their detection is mostly driven by Fe I, the authors constrain the abundances of all these species. For TiO they find an upper limit on the volume mixing ratio of 10$^{-8.5}$, which is inconsistent with the detection reported by \citet{Changeat2021}. 

\citet{Pino2022} also studied KELT-9b's Fe I emission with phase-resolved high-resolution spectra from HARPS-N and CARMENES.  Their analysis of the Fe I emission's Doppler shifts as a function of the orbital phase revealed a radial velocity anomaly that is likely explained by KELT-9b's atmospheric circulation patterns. Additionally, their analysis of the Fe I line depths shows them to be symmetric about the substellar point, indicating a lack of a hot-spot offset at the altitude probed by the Fe I emission lines.

To better understand KELT-9b's atmosphere, we observed partial phase curves using CARMENES\footnote{Some of these data were also used in \citet{Pino2022}.} and searched for emission from several species, including those previously detected in transmission.  Our observations found evidence for emission by Ca II, Mg I, and Si I and confirmed the previous detections of neutral iron emission.  

\section{Observations} \label{sec:data}

\begin{figure*}[!htb]
\centering
\includegraphics[width=\textwidth]{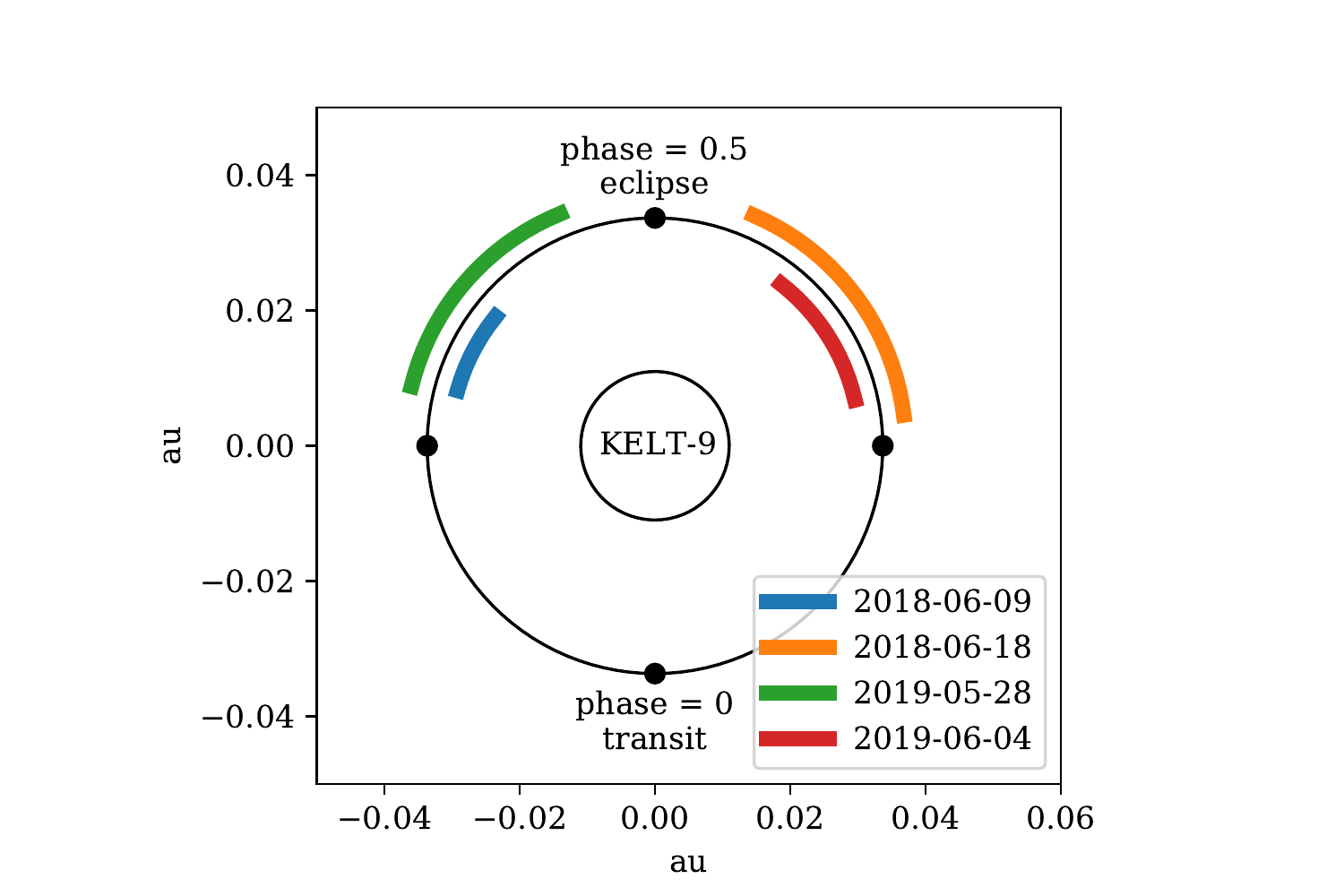} 
\caption{The phase coverage of our observed data sets.  The central black circle and the outer black circle are to scale representations of the radius of KELT-9 and the orbital semi-major axis of KELT-9b, respectively.}
\label{fig:PhaseCoverageDiagram}
\end{figure*}

We observed four partial phase curves of KELT-9b with the Calar Alto high-Resolution search for M dwarfs with Exoearths with Near-infrared and optical Echelle Spectrographs \citep[CARMENES;][]{Quirrenbach2014}, located at the Calar Alto Astronomical Observatory.  CARMENES utilizes a dichroic that splits the incoming light into separate visible (VIS) and near-infrared (NIR) channels.  The visible channel has a wavelength coverage of 520$-$960 nm at a resolving power of $R$ $\sim$ 94,600 and the near-infrared channel covers 960$-$1710 nm at a resolving power of $R$ $\sim$ 80,400.  The phase coverage of these observations is shown in Figure \ref{fig:PhaseCoverageDiagram} and more details are given in Table \ref{tab:ObsInfo}.  The first 10 frames of Night 3 (2019 May 28) have approximately three to four times higher noise than the subsequent frames so they were excluded from the analysis.  

\begin{table*}[!htb]
\caption{\label{tab:ObsInfo} Summary of Observations}
\begin{tabular}{ccccccc} \hline  \hline
Night & Date & Duration & Exposure & No. & Orbital & Average Signal-to-noise \\
& (UT) & (hr) & Time (s) & Frames & Phase & Ratio per Pixel$^\star$ \\ \hline
1 & 2018-06-09 & 2.30 & 112 & 38 & 0.64 $-$ 0.71 & 22.7 \\
2 & 2018-06-18 & 5.90 & 110 & 51 & 0.27 $-$ 0.44 & 27.5 \\
3 & 2019-05-28 & 5.34 & 172 & 85(75)$^\dagger$ & 0.56 $-$ 0.71 & 31.9 \\
4 & 2019-06-04 & 3.86 & 112 $-$ 172 & 43 & 0.29 $-$ 0.39 & 25.5 \\ \hline
\multicolumn{7}{l}{{$^\dagger$} The first 10 frames were excluded due to having very high noise.} \\
\multicolumn{7}{l}{{$^\star$} Estimated from the median standard deviation per order after applying \texttt{SYSREM}.} \\
\end{tabular}
\end{table*}

\section{Data Reduction} \label{sec:DataReduction}

The raw data were reduced with the \texttt{CARACAL} pipeline \cite{Caballero2016} at the observatory to produce one-dimensional wavelength calibrated spectra. The pipeline also provided a Barycentric Earth Radial Velocity correction for each frame.

Subsequent reduction steps were similar to previous work \citep[e.g.,][]{Herman2020,Turner2020,Deibert2021}. We corrected for cosmic rays by binning the data and removing points that were more than 5 median absolute deviations away from the median of the bin.  We also masked telluric emission lines according to \citet{Oliva2015}.  

Next we corrected for the variations caused by the instrument's blaze response function in each order separately.  To do this, we first divided every spectrum by the average spectrum of that order. Points in this division product lying more than 7.5 median absolute deviations from the median were removed, before fitting a third-order polynomial. Each spectrum was then blaze corrected by dividing by the evaluated polynomial fit.  An example of the blaze correction is shown in the top two panels of Figure \ref{fig:DataProcessingSteps}.  Finally, we masked wavelengths where at least one frame had a flux less than 10\% of the local continuum. This masking criterion resulted in orders 58 $-$ 61 being completely masked. 

\section{Correcting for Telluric Lines and Other Systematic Effects \label{sec:SYSREM}} 

In order to search for the planet's weak spectral signal, telluric lines, and any other systematic effects need to be corrected. To do this, we used the \texttt{SYSREM} algorithm \citep{Tamuz2005}, which is commonly used to remove stationary features in time \citep[e.g.,][]{Birkby2013, Birkby2017, Brogi2014, Ridden-Harper2016, Esteves2017, Hawker2018, Hoeijmakers2018a, Alonso-Floriano2019, Deibert2019, Deibert2021, Gibson2019, Ridden-Harper2019, Sanchez-Lopez2019, Herman2020, Jindal2020, Turner2020}.  To ensure that the telluric lines were stationary, we applied \texttt{SYSREM} to the data in the telluric rest frame. We used the airmass at each exposure as the initial estimate of a significant systematic to remove.    

The \texttt{SYSREM} algorithm operates in an iterative manner, with each consecutive iteration reducing the total variance of the data by a smaller amount. This means that most of the systematic variance is typically removed within a few iterations.

The \texttt{SYSREM} algorithm accounts for the uncertainty of each data point, and reduces to principle component analysis in the case of every data point having the same uncertainty \citep[e.g.,][]{Tamuz2005}. We estimated the uncertainties of the normalized data according to
\begin{equation}
\sigma = \frac{\sigma_t \otimes \sigma_\lambda}{\sigma_{t,\lambda}}
\end{equation}

where $\sigma_t$ and $\sigma_\lambda$ are the standard deviations in the time and wavelength dimensions, respectively, $\sigma_{t,\lambda}$ is the standard deviation in both the time and wavelength dimensions, and $\otimes$ is the outer product operator. 

\begin{figure*}[!htb]
\centering
\includegraphics[width=0.9\textwidth]{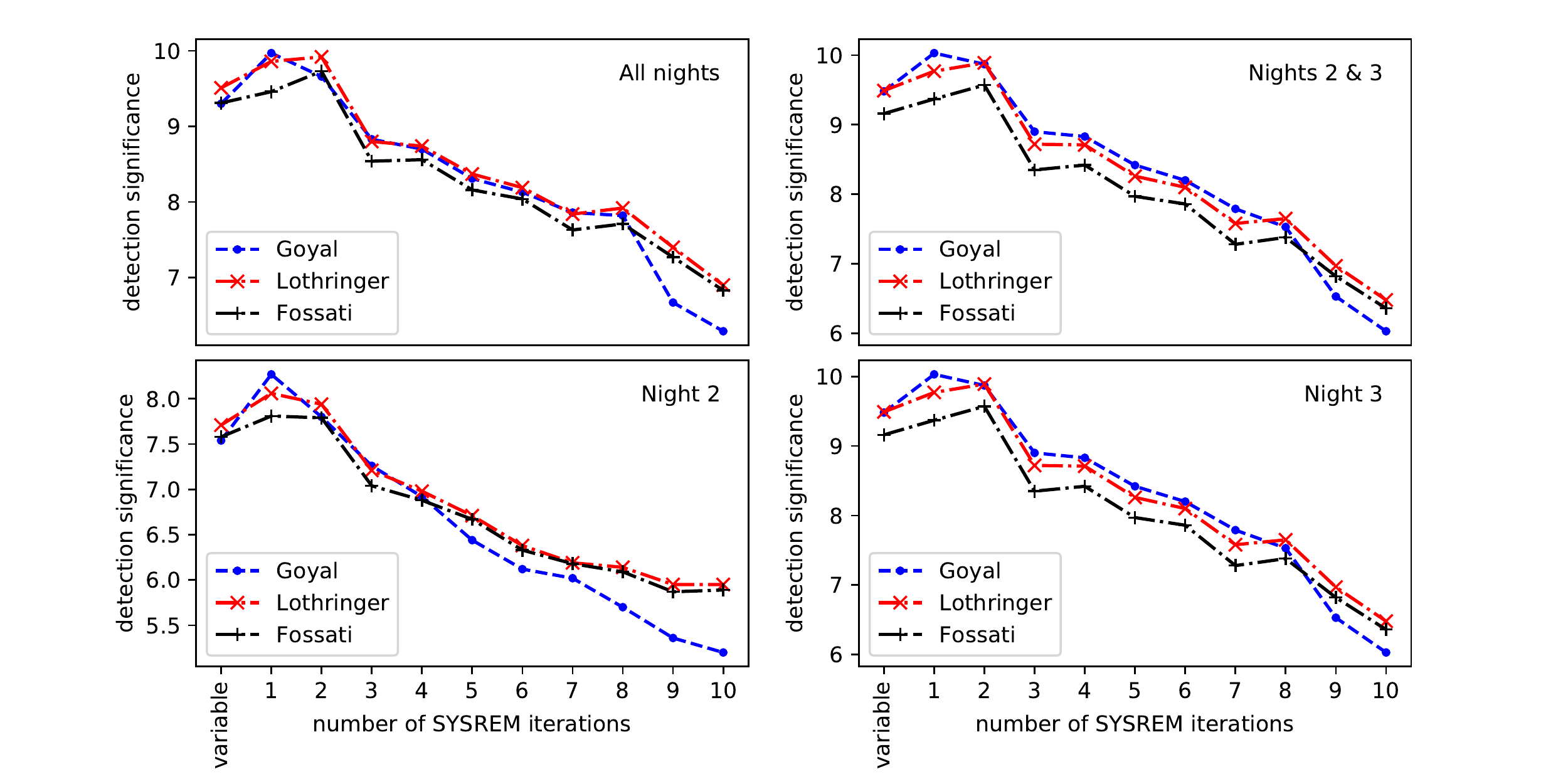} 
\caption{Detection significance of Fe I as a function of the number of \texttt{SYSREM} iterations applied to the data. The blue, red, and black lines use the \citet{Goyal2020},  \citet{Lothringer2018}, and \cite{Fossati2021} pressure-temperature profiles, respectively. The top left, top right, lower left, and lower right show the result of combining all nights, combining Nights 2 and 3, Night 2 alone, and Night 3 alone, respectively.}
\label{fig:FeDetSigWithSYSREM}
\end{figure*}

To determine the optimal number of \texttt{SYSREM} iterations to apply for each order, we considered (1) the variance of the data and (2) the detection significance of Fe I. For (1) we considered each order individually and examined how the variance of the data decreased as a function of the number of \texttt{SYSREM} iterations.    We found that a few iterations of \texttt{SYSREM} corrected the telluric lines and dominant systematics for the majority of orders. The number of iterations used on each order is shown in Table \ref{tab:SYSREM_iterations}. For (2), we applied a constant number of \texttt{SYSREM} iterations to all orders and examined how this affected the detection significance of Fe I (determined as in Section \ref{sec:DetectedSpecies}).  As shown in Fig. \ref{fig:FeDetSigWithSYSREM}, applying two \texttt{SYSREM} iterations to all orders resulted in the most robust Fe I detection so this was adopted for the rest of the analysis.

\begin{center}
\begin{longtable*}{c c c c c c}
\caption{The Number of SYSREM Iterations Used for Each Order and Night.} \label{tab:SYSREM_iterations} \\

\hline & Order & 2018-06-09 & 2018-06-18 & 2019-05-28 & 2019-06-04 \\ \hline 
\endfirsthead

\multicolumn{6}{c}
{{\bfseries \tablename\ \thetable{} -- continued from previous page}} \\
\hline  & Order & 2018-06-09 & 2018-06-18 & 2019-05-28 & 2019-06-04 \\ \hline 
\endhead

\hline \multicolumn{6}{|r|}{{Continued on next page}} \\ \hline
\endfoot

\hline \hline
\endlastfoot

        VIS & 118 & 1 & 2 & 2 & 5 \\ 
         & 117 & 1 & 2 & 4 & 4 \\ 
         & 116 & 2 & 1 & 2 & 2 \\ 
         & 115 & 3 & 1 & 2 & 2 \\ 
         & 114 & 3 & 3 & 2 & 2 \\ 
         & 113 & 5 & 1 & 3 & 2 \\ 
         & 112 & 4 & 1 & 2 & 2 \\ 
         & 111 & 1 & 2 & 3 & 2 \\ 
         & 110 & 5 & 2 & 1 & 2 \\ 
         & 109 & 3 & 2 & 2 & 2 \\ 
         & 108 & 2 & 2 & 3 & 3 \\ 
         & 107 & 3 & 1 & 3 & 2 \\ 
         & 106 & 1 & 1 & 3 & 2 \\ 
         & 105 & 3 & 2 & 3 & 2 \\ 
         & 104 & 1 & 2 & 3 & 2 \\ 
         & 103 & 4 & 3 & 3 & 2 \\ 
         & 102 & 2 & 2 & 2 & 2 \\ 
         & 101 & 4 & 2 & 1 & 2 \\ 
         & 100 & 4 & 1 & 1 & 2 \\ 
         & 99 & 4 & 2 & 1 & 2 \\ 
         & 98 & 4 & 1 & 2 & 2 \\ 
         & 97 & 2 & 2 & 2 & 3 \\ 
         & 96 & 3 & 1 & 1 & 2 \\ 
         & 95 & 3 & 1 & 2 & 2 \\ 
         & 94 & 2 & 1 & 2 & 1 \\ 
         & 93 & 2 & 1 & 3 & 1 \\ 
         & 92 & 1 & 1 & 1 & 1 \\ 
         & 91 & 2 & 1 & 1 & 1 \\ 
         & 90 & 1 & 1 & 1 & 1 \\ 
         & 89 & 3 & 1 & 2 & 2 \\
         & 88 & 3 & 1 & 2 & 3 \\ 
         & 87 & 2 & 1 & 1 & 2 \\ 
         & 86 & 1 & 1 & 1 & 2 \\ 
         & 85 & 2 & 1 & 2 & 2 \\ 
         & 84 & 2 & 1 & 3 & 2 \\ 
         & 83 & 2 & 2 & 3 & 2 \\ 
         & 82 & 1 & 1 & 2 & 2 \\ 
         & 81 & 1 & 3 & 4 & 2 \\ 
         & 80 & 1 & 2 & 5 & 2 \\ 
         & 79 & 1 & 1 & 1 & 3 \\ 
         & 78 & 1 & 1 & 1 & 2 \\ 
         & 77 & 1 & 1 & 2 & 3 \\ 
         & 76 & 1 & 2 & 3 & 3 \\ 
         & 75 & 3 & 3 & 3 & 4 \\ 
         & 74 & 2 & 3 & 3 & 4 \\ 
         & 73 & 2 & 3 & 3 & 2 \\ 
         & 72 & 2 & 1 & 3 & 2 \\ 
         & 71 & 2 & 1 & 3 & 2 \\ 
         & 70 & 1 & 1 & 2 & 2 \\ 
         & 69 & 2 & 2 & 3 & 2 \\ 
         & 68 & 3 & 3 & 3 & 3 \\ 
         & 67 & 4 & 2 & 4 & 4 \\ 
         & 66 & 2 & 3 & 3 & 5 \\ 
         & 65 & 1 & 4 & 3 & 5 \\ 
         & 64 & 1 & 2 & 4 & 5 \\ 
         & 63 & 1 & 3 & 2 & 5 \\ 
         & 62 & 1 & 3 & 3 & 2 \\ 
         & 61 & 2 & 3 & 3 & 2 \\ 
         & 60 & 3 & 3 & 3 & 2 \\ 
         & 59 & 2 & 3 & 3 & 4 \\ 
         & 58 & 2 & 3 & 3 & 5 \\ 
        NIR & 63 & 5 & 2 & 5 & 2 \\ 
         & 62 & 4 & 2 & 3 & 2 \\ 
         & 61 & 1 & 1 & 2 & 2 \\ 
         & 60 & 1 & 1 & 2 & 2 \\ 
         & 59 & 1 & 1 & 1 & 2 \\ 
         & 58 & 1 & 1 & 1 & 2 \\ 
         & 57 & 1 & 1 & 2 & 2 \\ 
         & 56 & 1 & 1 & 2 & 2 \\ 
         & 55 & 1 & 2 & 2 & 2 \\ 
         & 54 & 1 & 2 & 2 & 4 \\ 
         & 53 & 1 & 3 & 4 & 3 \\ 
         & 52 & 3 & 2 & 4 & 2 \\ 
         & 51 & 2 & 3 & 5 & 3 \\ 
         & 50 & 1 & 2 & 3 & 2 \\ 
         & 49 & 1 & 1 & 1 & 1 \\ 
         & 48 & 1 & 2 & 2 & 1 \\ 
         & 47 & 1 & 2 & 4 & 1 \\
         & 46 & 4 & 3 & 4 & 4 \\ 
         & 45 & 1 & 1 & 5 & 1 \\ 
         & 44 & 6 & 2 & 5 & 1 \\ 
         & 43 & 6 & 1 & 2 & 1 \\ 
         & 42 & 3 & 2 & 3 & 1 \\ 
         & 41 & 3 & 3 & 3 & 2 \\ 
         & 40 & 2 & 1 & 3 & 2 \\ 
         & 39 & 1 & 2 & 2 & 3 \\ 
         & 38 & 1 & 1 & 3 & 2 \\ 
         & 37 & 1 & 2 & 3 & 2 \\ 
         & 36 & 1 & 3 & 3 & 2 \\ \hline
         \multicolumn{6}{l}{Note. The orders are listed in order of increasing wavelength.}

\end{longtable*}
\end{center}

\begin{figure*}[!htb]
\centering
\includegraphics[width=0.80\textwidth]{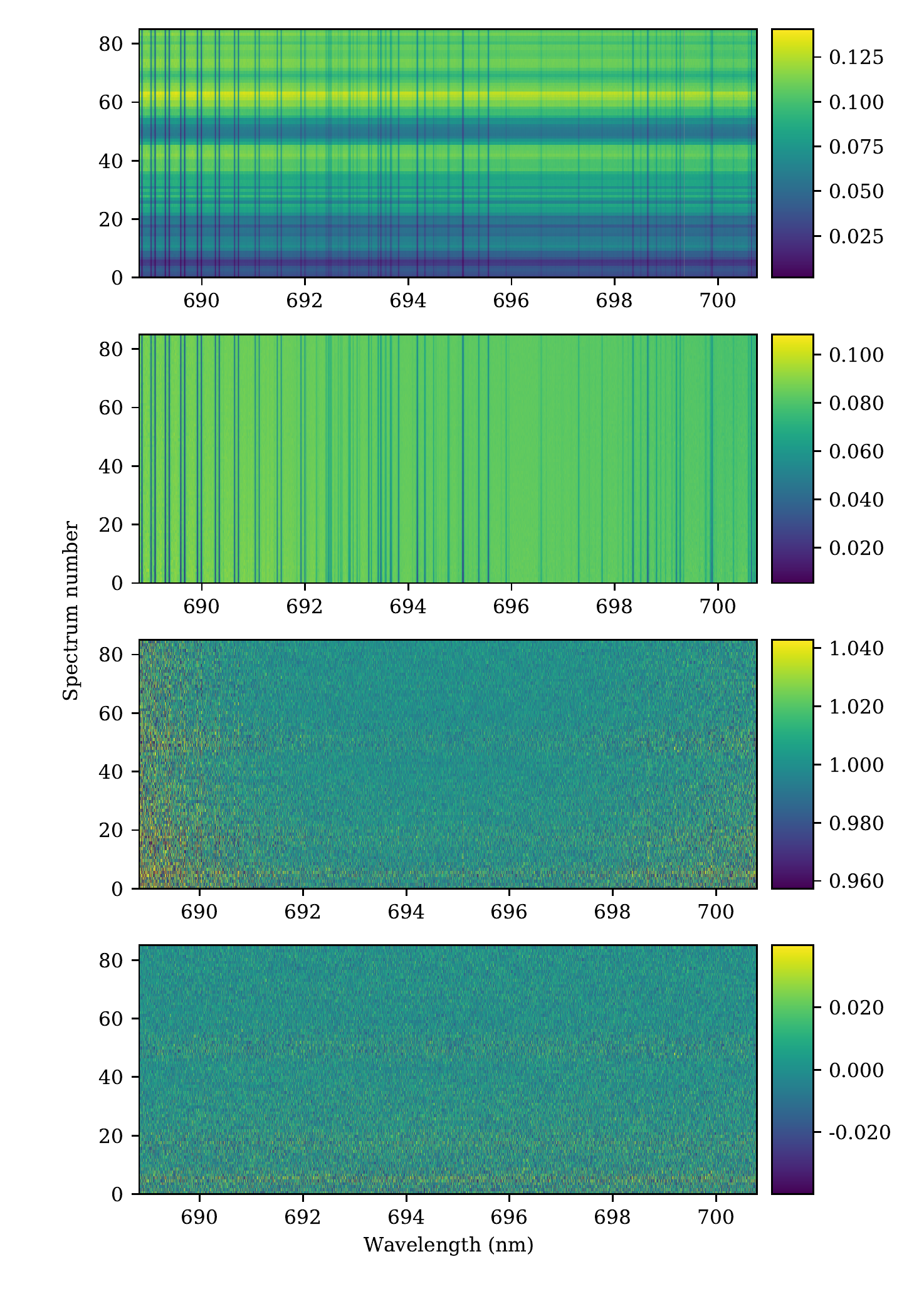} 
\caption{Summary of the data processing steps used in this analysis, shown for Night 3. The top panel shows the initial one-dimensional spectra from the \texttt{CARACAL} pipeline. The second panel shows the blaze-corrected spectra. The third panel shows the data after one \texttt{SYSREM} iteration. The fourth panel shows the data after weighing down wavelengths with high noise.}
\label{fig:DataProcessingSteps}
\end{figure*}

The \texttt{SYSREM} algorithm is less effective for orders with many strong and closely spaced telluric lines. To reduce the impact of these orders on our results, we weigh down each pixel by the variance over all frames at its wavelength with our cross-correlation implementation  \citep[see Equation \ref{eqn:CCF_definition}; e.g.,][]{Snellen2010,Ridden-Harper2016,Ridden-Harper2019,Herman2020,Deibert2021}.
The result of the \texttt{SYSREM} algorithm and the subsequent weighting for a typical order is shown in the third and fourth panels of Figure \ref{fig:DataProcessingSteps}.  Finally, we removed any outlying points more than five standard deviations away from the mean.

\section{Model spectra} \label{sec:model_spectra}

\begin{figure*}[!htb]
\centering
\includegraphics[width=0.80\textwidth]{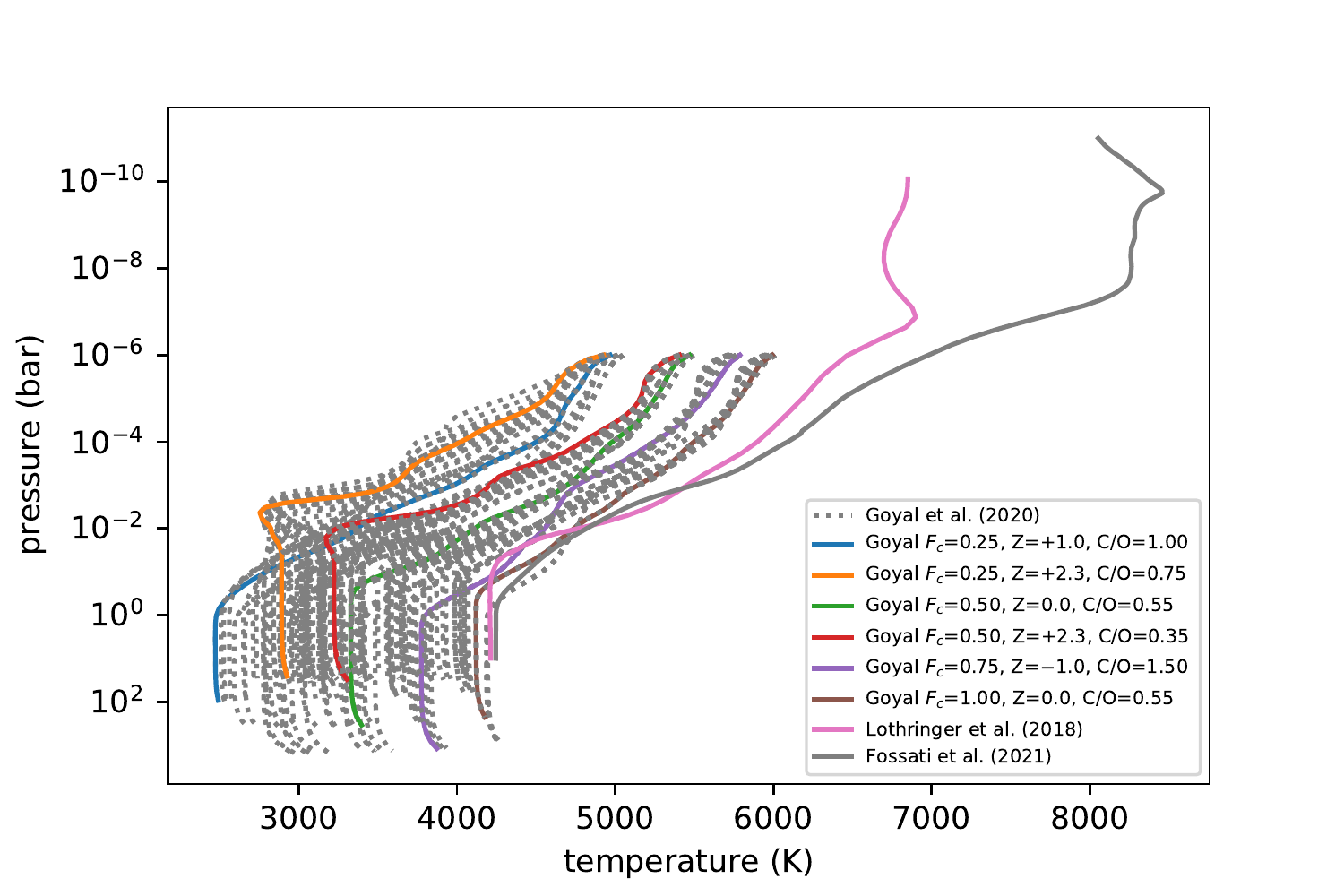} 
\caption{Pressure-temperature profiles used to create model emission spectra of KELT-9b. The profiles from \citet{Goyal2020} are only defined for pressures greater than 10$^{-6}$ bar.}
\label{fig:PT_profiles}
\end{figure*}

To search for species in the emission spectrum of KELT-9b with the cross-correlation method, we required model emission spectra. We used the publicly available software \texttt{petitRADTRANS}\footnote{https://petitradtrans.readthedocs.io/en/latest/} \citep{Molliere2019} to generate model emission spectra, based on given pressure-temperature ($P$-$T$) profiles and atmospheric abundances.  The sources of opacity data used by \texttt{petitRADTRANS} can be found in \citet{Molliere2019}.  Notably, it includes the recent ``TOTO'' linelist for TiO \citep{McKemmish2019}. We adopted KELT-9b $P$-$T$ profiles from three sources: \citet{Goyal2020}, \citet{Lothringer2018}, and \citet{Fossati2021}.  These profiles are shown in Fig. \ref{fig:PT_profiles}.

The KELT-9b atmospheric model adopted from  \citet{Goyal2020} is part of a library of 89 self-consistent simulated exoplanet atmospheres. For each planet, they simulated a grid of model atmospheres (including $P$-$T$ profiles and abundances) that spans the parameter space of recirculation factor $f_c$ = 0.25,  0.5,  0.75,  1.0, metallicity = 0.1,  1,  10,  50,  100, 200 times solar, and C/O ratio = 0.35, 0.55, 0.7, 0.75 1.0, 1.5. The $P$-$T$ profiles and abundances of the 277 chemical species used in their equilibrium chemistry calculations are publicly available.\footnote{\url{https://drive.google.com/drive/folders/1g7Bc6pbwvLUDf-QFJOCEOFKX6glKLuqF}}  As the models of \citet{Goyal2020} were designed for traditional hot Jupiters, they lack opacity sources that are important for an ultrahot Jupiter like KELT-9b. Therefore, we also used the $P$-$T$ profiles of \citet{Lothringer2018} and \citet{Fossati2021}, which were tailored to KELT-9b. We used the publicly available software \texttt{GGchem}\footnote{\url{https://github.com/pw31/GGchem}} \citep{Woitke2018} to generate abundance profiles for a given $P$-$T$ profile. These abundance profiles assumed a solar atmospheric composition, and no condensation.

The abundances provided by \citet{Goyal2020} and \texttt{GGchem} were in mole fractions and particle densities, respectively. Therefore, we first converted them into mass fractions for use by \texttt{petitRADTRANS}.  As the \citet{Goyal2020} $P$-$T$ profiles span a smaller pressure range than those of \citet{Lothringer2018} and \citet{Fossati2021}, we interpolated them onto different pressure grids for use by \texttt{petitRADTRANS}. For the \citet{Goyal2020} models, we used a pressure grid consisting of 130 pressure levels evenly spaced on a log scale that spanned the range of $10^{-6} - 10^3$ bar.  For the \citet{Lothringer2018} and \citet{Fossati2021} models, we used 200 pressure levels evenly spaced on log scales that spanned the ranges of $10^{-10} - 10^1$ bar and $10^{-11} - 10^1$ bar, respectively. In order to search for the emission lines of specific species, we created models that included line-by-line opacities for only one species. However, each of our models also included Rayleigh scattering cross sections of H$_2$, He, N$_2$ and continuum opacities of H$_2$-H$_2$, H$_2$-He, and H$^-$.  We created models with only a single species' emission lines for Al I, Ca I, Ca II, Cr I, Fe I, FeH, Fe II, K I, Li I, Mg I, Mg II, Na I, OH, Si I, Ti I, TiO, V I, V II, VO, Y I.  We omitted Cr II, Sc II, Y II, and Ti II, which were seen in transmission \citep{Hoeijmakers2019}, as their opacities were not included in \texttt{petitRADTRANS}. Following \citet{Pino2020}, we neglected the contribution of reflected stellar light to KELT-9b's emission spectrum for two main reasons. The first reason is that phase curve observations indicate that KELT-9b has a low albedo \citep{Jones2022}. The second reason is KELT-9's high rotation \citep[111.4 $\pm$ 1.3 km s$^{-1}$; ][]{Gaudi2017} results in broad spectral lines that are difficult to detect with high-resolution cross-correlation spectroscopy. The insensitivity of  high-resolution cross-correlation spectroscopy to broad stellar lines was demonstrated by \citet{Spring2022} who were unable to detect reflected light from 51 Pegasi b.

\texttt{petitRADTRANS} generates model emission spectra at a resolving power of $R$ = 10$^6$ at all wavelengths.  Before downgrading this resolution, we applied a rotational broadening of 6.63 km s$^{-1}$, corresponding to KELT-9b being tidally locked \citep{Pino2020}.  We carried out this broadening with the rotational broadening function in the \texttt{Eniric} software\footnote{\url{https://eniric.readthedocs.io/en/latest/index.html}} \citep{Figueira2016}.

After rotationally broadening the spectra, we downgraded their spectral resolution to correspond to the resolving power of CARMENES ($R$ $\sim$ 94,600 in the visible channel and $R$ $\sim$ 80,400 in the NIR channel) by resampling them onto new wavelength grids at the desired resolution at all wavelengths.  We used the \texttt{SpectRes} software\footnote{\url{https://spectres.readthedocs.io/en/latest/}} \citep{Carnall2017} for this resampling, which preserves integrated flux.

By performing a blaze correction (see Section \ref{sec:DataReduction}) and applying \texttt{SYSREM} (see Section \ref{sec:SYSREM}), we effectively divide the data by the time-averaged emission spectrum of the KELT-9 system. As KELT-9's emission is much stronger than KELT-9b's, the time-averaged emission of the system can be well approximated as the emission of only KELT-9.  Following, e.g. \citet{Nugroho2021} and \citet{Herman2022}  we therefore scaled our KELT-9b emission models before using them as cross-correlation templates according to

\begin{equation}
    F = \left( \frac{R_p}{R_s} \right)^2 \frac{F_p}{F_s}
\end{equation}

where $R_p$ is the radius of KELT-9b, $R_s$ is the radius of KELT-9, $F_p$ is the emission spectrum of KELT-9b (provided by \texttt{petitRADTRANS}) and $F_s$ is the spectrum of KELT-9, which we approximated as a blackbody with an effective temperature of 10170 K.  

\begin{figure*}[!htb]
\centering
\includegraphics[width=0.85\textwidth]{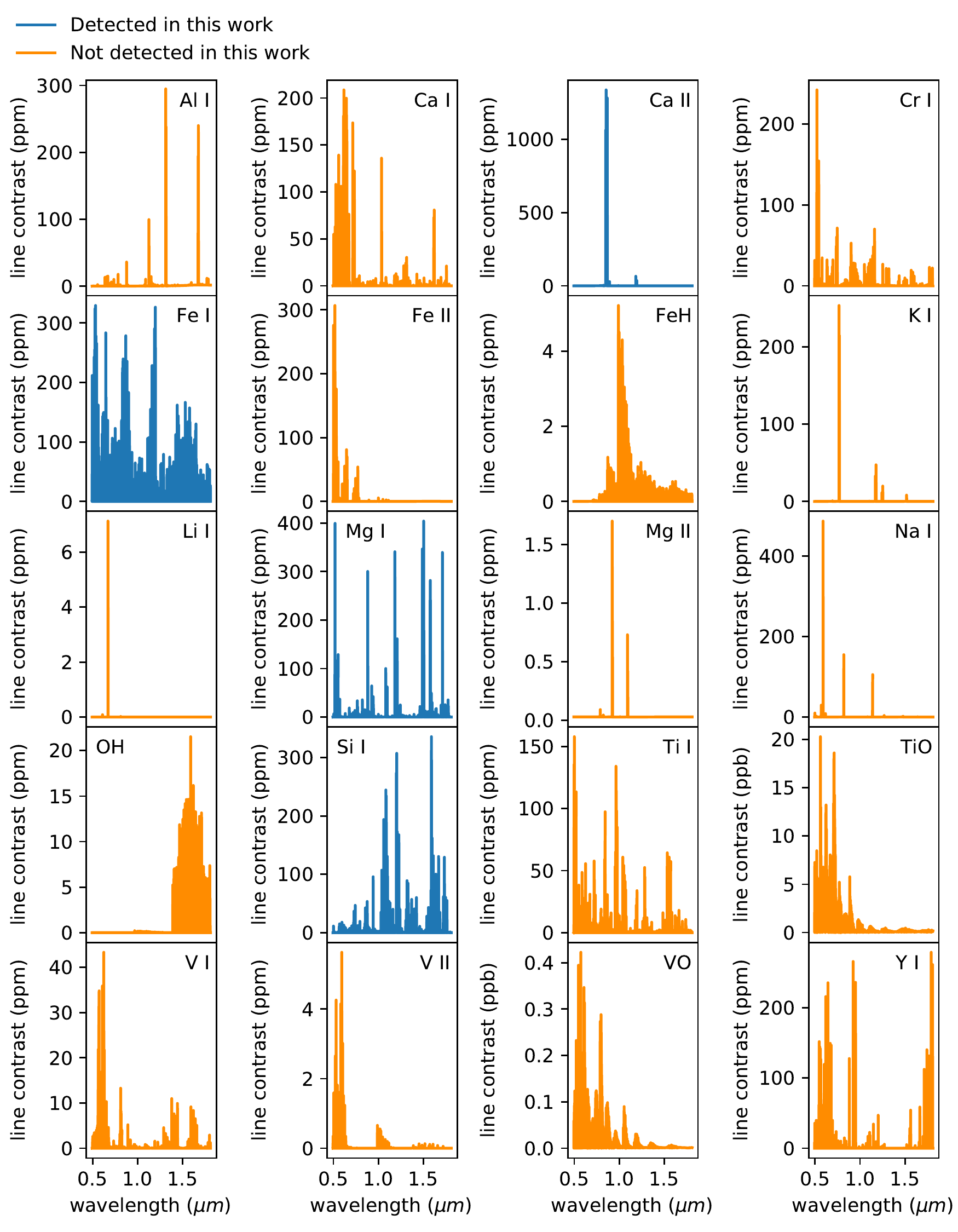} 
\caption{Continuum-subtracted model emission spectra of individual species in the atmosphere of KELT-9b, assuming the \citet{Lothringer2018} pressure-temperature profile. We find evidence for the presence of species shown in blue but no evidence for those shown in orange. The vertical axes for all species are in units of parts per million (ppm), except for TiO and VO, which are in units of parts per billion (ppb).}
\label{fig:ModelSpectra}
\end{figure*}

The cross-correlation method used in this analysis is sensitive to line contrasts above the broadband continuum of KELT-9b's spectrum. However, the model emission spectra generated by \texttt{petitRADTRANS} include the continuum so we fit a polynomial to the continuum and subtracted it from the model spectra.  Some of the continuum-subtracted model emission spectra used in this analysis are shown in Figure \ref{fig:ModelSpectra}.   

\section{Cross Correlation} \label{sec:analysis}

After reducing the data and correcting for telluric lines and systematics, we searched for atmospheric species with the cross-correlation method. Each model was cross-correlated with one corrected spectrum at a time to produce a cross-correlation function, $\mathrm{CCF}$, calculated according to 

\begin{equation}
\label{eqn:CCF_definition}
    \textrm{CCF}(v_{sys}) = \sum_i \frac{f_i m_i(v_{sys})}{\sigma_i^2}.
\end{equation}

where $f_i$ is the mean-subtracted spectrum, $m_i$ is the mean-subtracted Doppler-shifted model, shifted to velocity $v_{sys}$, and $\sigma$ is the uncertainty per data point as defined in Section \ref{sec:SYSREM}. We varied $v_{sys}$ from $-$1000 to $+$1000 km s$^{-1}$ in steps of 1 km s$^{-1}$.

We corrected for any offsets in our CCFs by subtracting the median value from each CCF in the velocity direction before summing CCFs across orders to form a single CCF per frame per night.

To combine the summed CCFs of each night, we shifted them into KELT-9b's rest frame. We calculated the phase, $\phi(t)$, of KELT-9b at the time of each exposure using the ephemeris given in Table \ref{tab:PlanetParameters} and the corresponding radial velocity, $v_r$, as  $v_r = K_p \sin(2 \pi \phi(t))$ where $K_p = v_{orb} \sin(i)$ and $v_{orb}$ and $i$ are KELT-9b's orbital velocity and inclination, respectively.  At this stage, we also corrected for the Earth's barycentric radial velocity using the values provided by the CARACAL pipeline.  Once aligned in KELT-9b's rest frame, the CCFs in each order were combined by averaging.

\begin{table*}[!htb]
\centering
\caption{\label{tab:PlanetParameters} Parameters of KELT-9b}

\begin{tabular}{lll} \hline  \hline
Parameter & Value & Reference  \\ \hline
Orbital period (days) & 1.4811235 & \citet{Ahlers2020}   \\
Mid-transit (BJD) & 2457095.68572 & \citet{Ahlers2020}   \\
Semimajor axis (au) & 0.035 & \citet{Gaudi2017}   \\
Inclination & 87.2  &  \citet{Ahlers2020}  \\
$R_P$ ($R_{\textrm{Jup}}$) & 1.891 & \citet{Gaudi2017} \\
$v_{sys}$ (km s$^{-1}$) & -20.5  & \citet{Pino2020}  \\
$K_p$  & 241.5 & \citet{Pino2020} \\ 
Star &  &  \\
$T_{\textrm{eff}}$ (K) & 10170  & \citet{Gaudi2017}   \\
Radius ($R_\odot$) & 2.36 &  \citet{Gaudi2017}  \\ 
Rotation rate (km s$^{-1}$) & 111.4 & \citet{Gaudi2017} \\ 

\hline
\end{tabular}

\end{table*}

We estimated the significance of potential detections from this cross-correlation search by dividing the amplitude of the phase-folded CCF at the expected $v_{sys}$ by the standard deviation of the surrounding phase-folded CCF.

To allow for the uncertainties in the literature value of $K_p$ and to explore the properties of the noise, we created cross-correlation maps that spanned a range of values for $K_p$ and systemic velocity ($v_{sys})$.  We varied $K_p$ from 0 to 300 km s$^{-1}$ in steps of 1 km s$^{-1}$.

To investigate whether coincidental similarities between models could lead to signals being misidentified, we cross-correlated all model emission spectra with each other and searched for signals within $\pm$10 km s$^{-1}$.  We focused on this velocity range as it includes the Doppler shifts from winds in some exoplanet atmospheres \citep[e.g.,][]{Snellen2010, Louden2015, Wyttenbach2015, Brogi2016}.  We found no cross-correlation signals between our models at 0 km s$^{-1}$ so they cannot produce a false signal at the expected velocity. However, we did find some cross-correlation signals at non-zero velocities that could be incorrectly attributed to atmospheric winds.  To further investigate this scenario, we also considered the cross-correlation amplitude relative to that of the autocorrelation of the shifted model. Our results for models using both the \citet{Lothringer2018} and \citet{Fossati2021} $P$-$T$ profiles are shown in Table \ref{tab:CCFsBetweenModels}. Most signals have cross-correlation amplitudes of $<$10\% of their corresponding autocorrelations, likely making them too weak to detect in noisy data. However, the relative amplitudes for the combinations of Fe I and Ca II, V I and Ca II, Ca I and Fe I, and Fe II and Fe I are between 10\% and 30\% so they have a higher potential for producing false signals mistakenly attributed to winds.

\begin{table*}[!htb]
\centering
\caption{Cross-correlation Signals Arising from Coincidental Model Similarities \label{tab:CCFsBetweenModels}}
\begin{tabular}{cccccc}
\hline \hline
              &              &  \multicolumn{2}{c}{\citet{Lothringer2018}}  & \multicolumn{2}{c}{\citet{Fossati2021}} \\
Shifted & Static & Velocity   & Amplitude Relative & Velocity & Amplitude Relative \\  
Model & Model & (km s$^{-1}$)  & to Auto-CCF & (km s$^{-1}$) & to Auto-CCF  \\ \hline
Ca II         & Fe I           & $-$9.5 & 3.7$\times 10^{-2}$ & $-$9.5 & 2.8$\times 10^{-2}$ \\
Fe I            & Ca II        & $+$9.5 & 6.7$\times 10^{-2}$ & $+$9.5 & 1.1$\times 10^{-1}$ \\
Ca II         & V I            & $-$3.2 & 1.0$\times 10^{-4}$ & $-$3.2 & 7.1$\times 10^{-5}$ \\
V I             & Ca II        & $+$3.2 & 5.1$\times 10^{-2}$ & $+$3.2 & 1.3$\times 10^{-1}$ \\
Fe I            & Ca I           & $-$6.3 & 1.7$\times 10^{-2}$ & $-$6.3 & 2.0$\times 10^{-2}$ \\
Ca I            & Fe I           & $+$6.4 & 1.7$\times 10^{-1}$ & $+$6.4 & 2.2$\times 10^{-1}$ \\
K I             & Cr I           & $-$7.4 & 2.9$\times 10^{-3}$ & $-$7.4 & 1.1$\times 10^{-3}$ \\
Cr I            & K I            & $+$7.5 & 1.0$\times 10^{-3}$ & $+$7.5 & 4.2$\times 10^{-4}$ \\
Fe II         & Fe I           & $-$6.3 & 3.3$\times 10^{-1}$ & $-$6.3 & 3.0$\times 10^{-1}$ \\
Fe I            & Fe II        & $+$6.4 & 2.1$\times 10^{-2}$ & $+$6.4 & 3.6$\times 10^{-2}$ \\
Y I             & Fe I           & $-$6.3 & 4.4$\times 10^{-2}$ & $-$6.3 & 3.9$\times 10^{-2}$ \\
Fe I            & Y I            & $+$6.4 & 6.2$\times 10^{-3}$ & $+$6.4 & 8.3$\times 10^{-3}$ \\ \hline             
\end{tabular}
\end{table*}

\section{Likelihood Mapping \label{sec:lnL}}

To robustly combine the four nights of data and constrain the properties of KELT-9b's atmosphere, we used the likelihood mapping method \citep{Brogi2019,Gibson2020}. This method extends the cross-correlation method described in Section \ref{sec:analysis}, making it possible to statistically quantify how well different models explain the data.  To account for phase variations, we follow the approach taken by \citet{Herman2022}, and adapt their publicly available Python code\footnote{\url{https://github.com/mkherman/HRS-logL}} for our data.  The approach is based on the formulation of \cite{Gibson2020} that generalizes the \cite{Brogi2019} likelihood to include both time- and wavelength-dependent uncertainties.  This can alternatively be seen as a version of the full likelihood of \cite{Gibson2020} that optimizes a global scale factor for the uncertainties.  The likelihood, $\mathcal{L}$, is calculated according to 

\begin{equation}
    \mathrm{ln} \mathcal{L} = -\frac{N}{2} \mathrm{ln} \frac{\chi^2}{N}
\end{equation}

where $N$ is the total number of data points and $\chi^2$ is related to the cross-correlation function, $\mathrm{CCF}$ as

\begin{equation} \label{eqn:chi2}
    \chi^2 = \sum \frac{f_i^2}{\sigma_i^2} + A_p(\phi)^2 \sum \frac{m_i^2}{\sigma_i^2} - 2A_p(\phi) \mathrm{CCF}
\end{equation}

where $f_i$ , $m_i$, $\sigma_i$, and $\textrm{CCF}$ are defined in Section \ref{sec:analysis}.  $A_p(\phi)$ is a phase-dependent term introduced by \cite{Herman2022} to model the planet's line contrast variation during the observations. It is defined as 

\begin{equation} \label{eqn:Ap}
    A_p(\phi) = \alpha(1-C \cos^2(\pi(\phi-\theta)))
\end{equation}

where $\alpha$ is a scaling factor to allow uncertainty in the scale of the model, $\theta$ is the phase offset of the peak line contrast, defined so that a positive offset occurs after the secondary eclipse.  $C$ is the day-night contrast given by $C = 1-F_n/F_d$ where $F_n$ and $F_d$ are the planetary line emissions of the day- and nightsides, respectively, measured over the wavelength range of our observations, i.e., $C$ = 0: no variation between day and nightside; $C$ = 1: no emission from the nightside.

The log-likelihoods are computed for each order, then summed together. To convert the log-likelihood map into likelihood, we subtracted the maximum value (occurring at the planet's $K_p$ and $v_{sys}$) to normalize, then took the exponential.  We then constructed 2D conditional distributions and 1D marginalized distributions to estimate the best-fit values and their uncertainties. We first used likelihood mapping to robustly fit the values of $K_p$ and $v_{sys}$ of detections identified by the classical cross-correlation search (see Section \ref{sec:analysis}). To do this, we only allowed $K_p$, $v_{sys}$, and $\alpha$ to vary, while the $C$ and $\theta$ were both fixed to zero. We varied the parameter grids' range and step sizes to ensure that the likelihood space was well-sampled in all cases. These parameter grids are shown in Table \ref{tab:lnLGrids}.  For the grid of models produced by \citet{Goyal2020},  we computed the log-likelihoods for each of their four recirculation factors, six metallicities, and six C/O ratios.

We also assessed how well our models match the data by using the Bayesian information criterion (BIC), defined as

\begin{equation}
    \mathrm{BIC} = k \mathrm{ln}(N) - 2 \mathrm{ln}(L_{max})
\end{equation}

where $\mathrm{ln}(L_{max})$ is the maximum log-likelihood and $k$ is the number of parameters that were constrained explicitly in the fits (e.g., $\alpha$, $K_p$, $v_{sys}$).

%%%%%%%%%%%%%%%%%%

\begin{table*}[!htb]
\centering
\caption{Parameter Grids Used for Log-Likelihood Analysis \label{tab:lnLGrids}}
\begin{tabular}{ccccc}
\hline \hline
Model                        & Parameter               & Start                          & End    & Step Size \\ \hline
Fe I                           & $\alpha$                & 0.00                           & 2.05   & 0.05      \\
\citet{Goyal2020}            & $K_p$ (km s$^{-1}$)     & 200                            & 300    & 1         \\
($f_c$=0.50, Z=0.0, C/O=0.5) & $v_{sys}$ (km s$^{-1}$) & -58                            & 7      & 1         \\
                             &                         &                                &        &           \\
Fe I                           & $\alpha$                & 0.00                           & 7.05   & 0.05      \\
\citet{Lothringer2018} and    & $K_p$ (km s$^{-1}$)     & 200                            & 300    & 1         \\
\citet{Fossati2021}          & $v_{sys}$ (km s$^{-1}$) & -58                            & 7      & 1         \\
                             &                         &                                &        &           \\
Ca II                        & $\alpha$                & 0.00                           & 3      & 0.05      \\
\citet{Lothringer2018} and    & $K_p$ (km s$^{-1}$)     & 100                            & 400    & 1         \\
\citet{Fossati2021}          & $v_{sys}$ (km s$^{-1}$) & -120                           & 120    & 1         \\
                             &                         &                                &        &           \\
Mg I                           & $\alpha$                & 0.0                            & 14.0   & 0.2       \\
\citet{Lothringer2018} and    & $K_p$ (km s$^{-1}$)     & 150                            & 320    & 1         \\
\citet{Fossati2021}          & $v_{sys}$ (km s$^{-1}$) & -70                            & 20     & 1         \\
                             &                         &                                &        &           \\
Si I                           & $\alpha$                & 0.0                            & 14.0   & 0.2       \\
\citet{Lothringer2018} and    & $K_p$ (km s$^{-1}$)     & 150                            & 400    & 1         \\
\citet{Fossati2021}          & $v_{sys}$ (km s$^{-1}$) & -70                            & 20     & 1         \\ \hline
Fe I                           & $\alpha$                & 0.0                            & 5.0    & 0.2       \\
\citet{Goyal2020}            & $K_p$ (km s$^{-1}$)     & 205                            & 285    & 1         \\
($f_c$=0.50, Z=0.0, C/O=0.5) & $v_{sys}$ (km s$^{-1}$) & -58                            & 7      & 1         \\
                             & $\theta$                & -180                           & 180    & 30        \\
                             & $C$                     & 0.00                           & 1.00   & 0.25      \\
                             &                         &                                &        &           \\
Fe I                           & $\alpha$                & 0.0                            & 14.0   & 0.5       \\
\citet{Lothringer2018} and    & $K_p$ (km s$^{-1}$)     & 205                            & 285    & 1         \\
\citet{Fossati2021}          & $v_{sys}$ (km s$^{-1}$) & -58                            & 7      & 1         \\
                             & $\theta$                & -180                           & 180    & 30        \\
                             & $C$                     & 0.0                            & 1.0    & 0.25      \\ \hline
Fe I                           & $\alpha$                & 0                              & 4.5    & 0.05      \\ 
\citet{Goyal2020}            & $K_p$ (km s$^{-1}$)     & 215                            & 265    & 1         \\
                             & $v_{sys}$ (km s$^{-1}$) & 48                             & 7      & 1         \\
                             & $\theta$                & -180                           & 180    & 30        \\
                             & $C$                     & 0.0                            & 1.0    & 0.25      \\
                             & $f_c$                   & 0.25                           & 1.00   & 0.25      \\
                             &                         &          \multicolumn{3}{c}{values}      \\
                             & log$_{10}$(Z)           & \multicolumn{3}{c}{-1.0, 0.0, 1,.0 1.7, 2.0, 2.3}     \\
                             & C/O                     & \multicolumn{3}{c}{0.35, 0.55, 0.7, 0.75 1.0, 1.5}   \\ \hline    
\end{tabular}
\end{table*}

%%%%%%%%%%%%%%%%%%%%

\section{Results and Discussion} \label{sec:results}

%%%%%%%%%%%%%%%%%%%%%%%%%%%%%%%
We searched for Al I, Ca I, Ca II, Cr I, Fe I, Fe II, FeH, K I, Li I, Mg I, Mg II, Na I, OH, Si I, Ti I, and TiO, V I, V II, VO, and Y I using models generated with the \citet{Lothringer2018} and \citet{Fossati2021} $P$-$T$ profiles.  We also searched for Fe I using the \citet{Goyal2020} model based on  solar metallicity, solar C/O ratio, and a recirculation factor of $f_c =$ 0.5\footnote{consistent with the value derived by \cite{Wong2020,Wong2021} from TESS phase curve observations.}.

\subsection{Detected Species \label{sec:DetectedSpecies}}

\begin{figure*}[!htb]
\centering
\begin{tabular}{lll}
  Fe I &  & \\
 \includegraphics[width=0.30\textwidth,page=1]{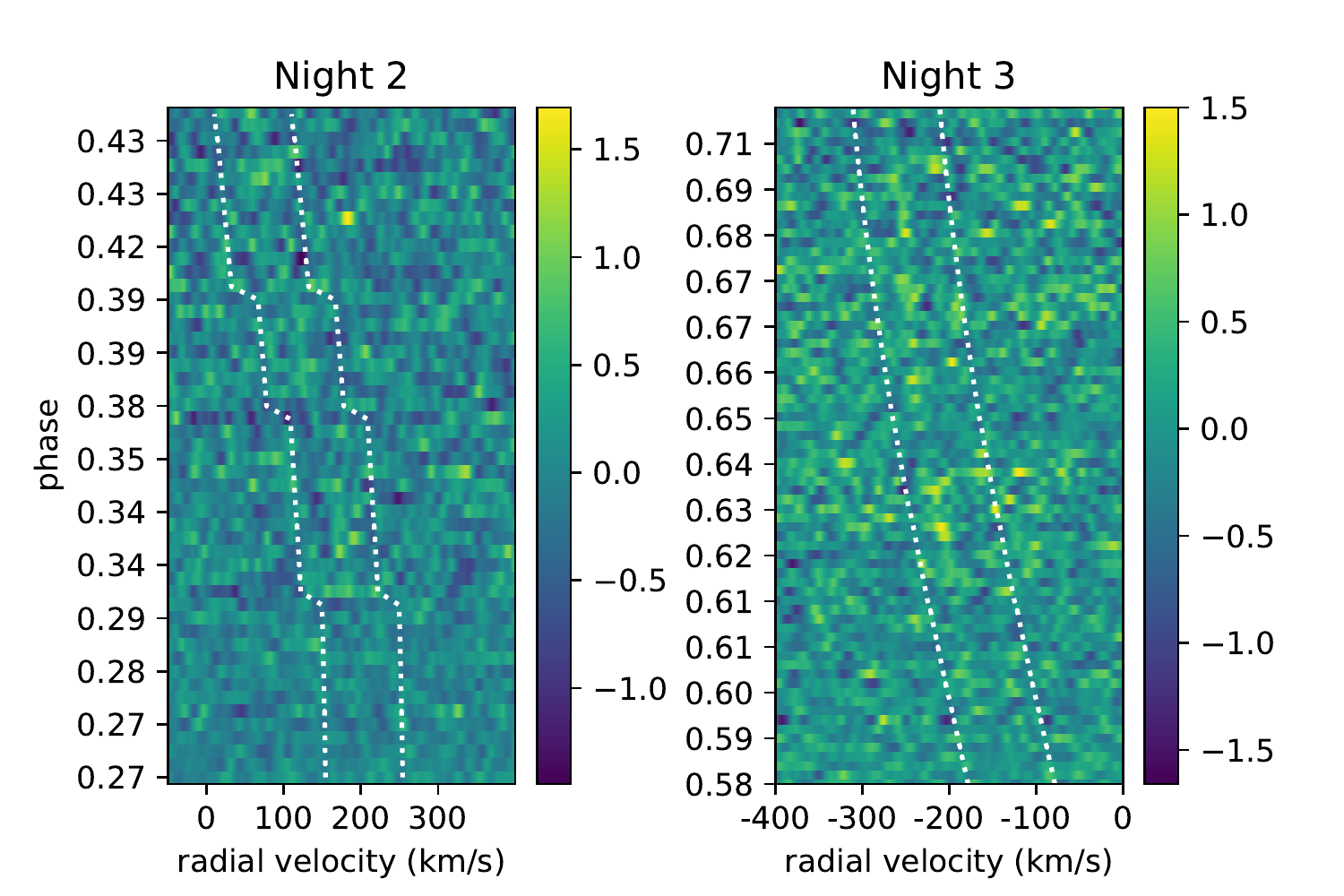} & 
 \includegraphics[width=0.30\textwidth,page=2]{Fe_Lothringer2018_Solar_for_paper.pdf} & 
  \includegraphics[width=0.30\textwidth,page=3]{Fe_Lothringer2018_Solar_for_paper.pdf} \\
    Ca II &  & \\
 \includegraphics[width=0.30\textwidth,page=1]{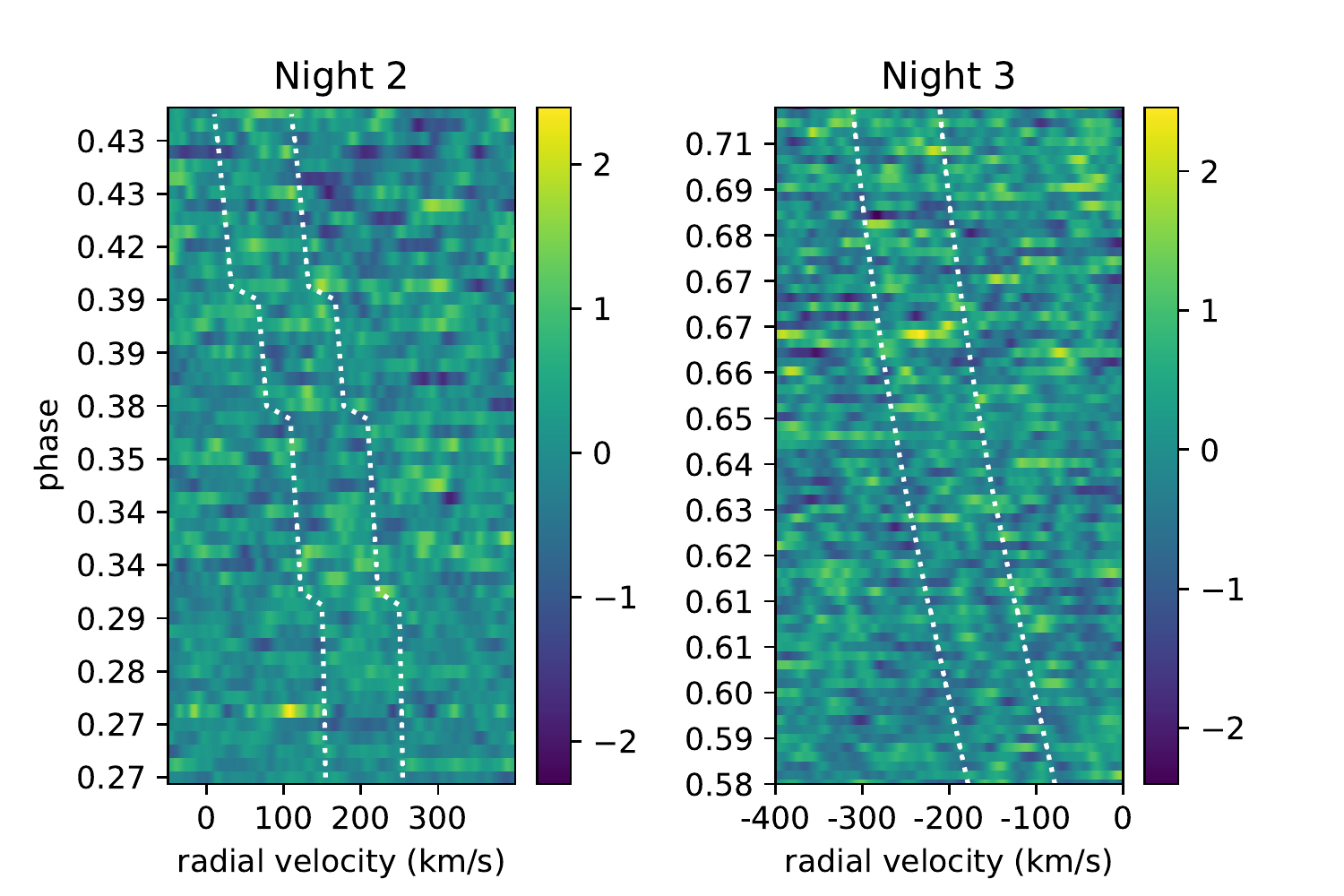} & 
 \includegraphics[width=0.30\textwidth,page=2]{CaII_Fossati2021_Solar_for_paper.pdf} & 
  \includegraphics[width=0.30\textwidth,page=3]{CaII_Fossati2021_Solar_for_paper.pdf} \\
  Mg I &  & \\
 \includegraphics[width=0.30\textwidth,page=1]{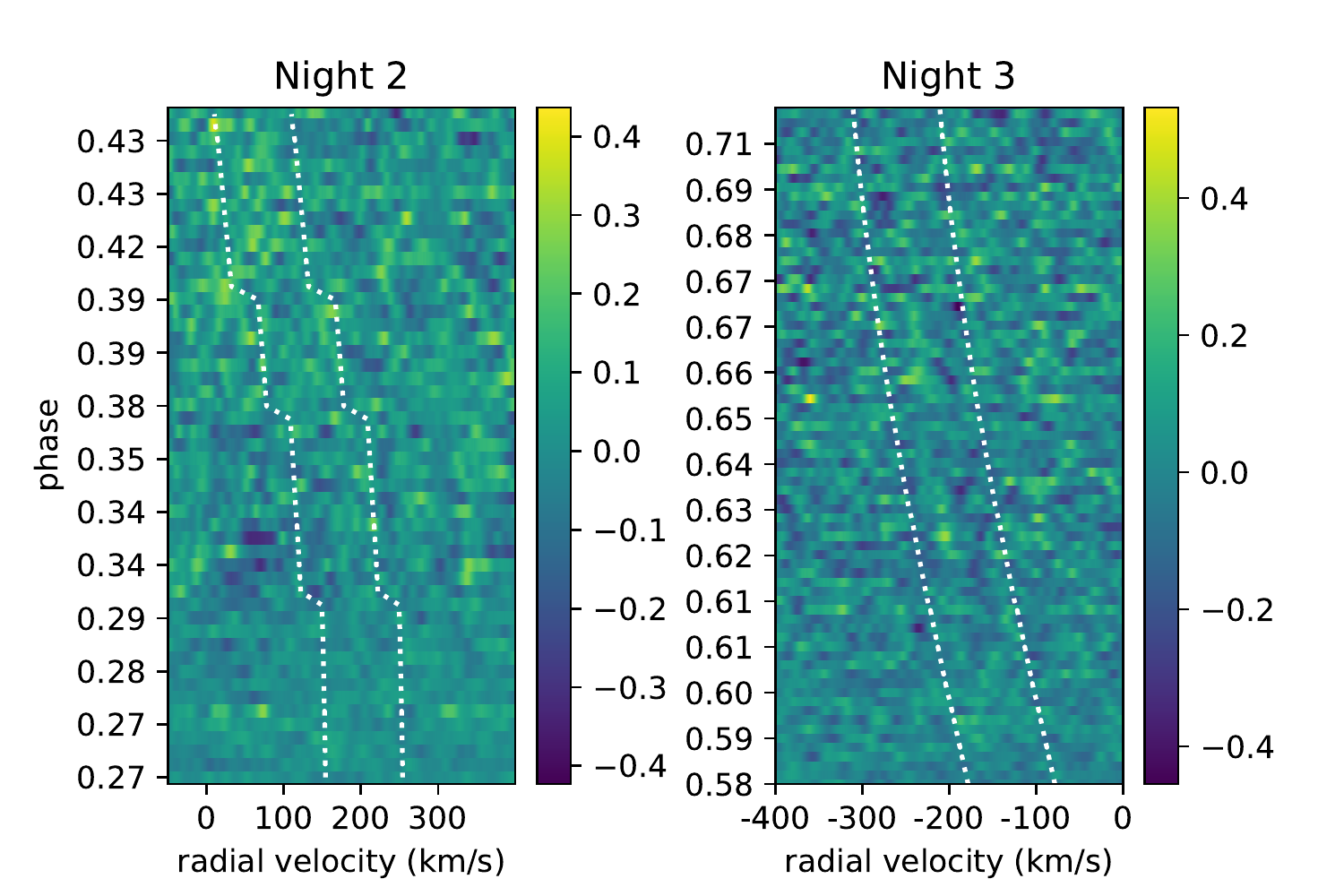} & 
 \includegraphics[width=0.30\textwidth,page=2]{Mg_Fossati2021_Solar_for_paper.pdf} & 
  \includegraphics[width=0.30\textwidth,page=3]{Mg_Fossati2021_Solar_for_paper.pdf} \\
  Si I &  & \\
 \includegraphics[width=0.30\textwidth,page=1]{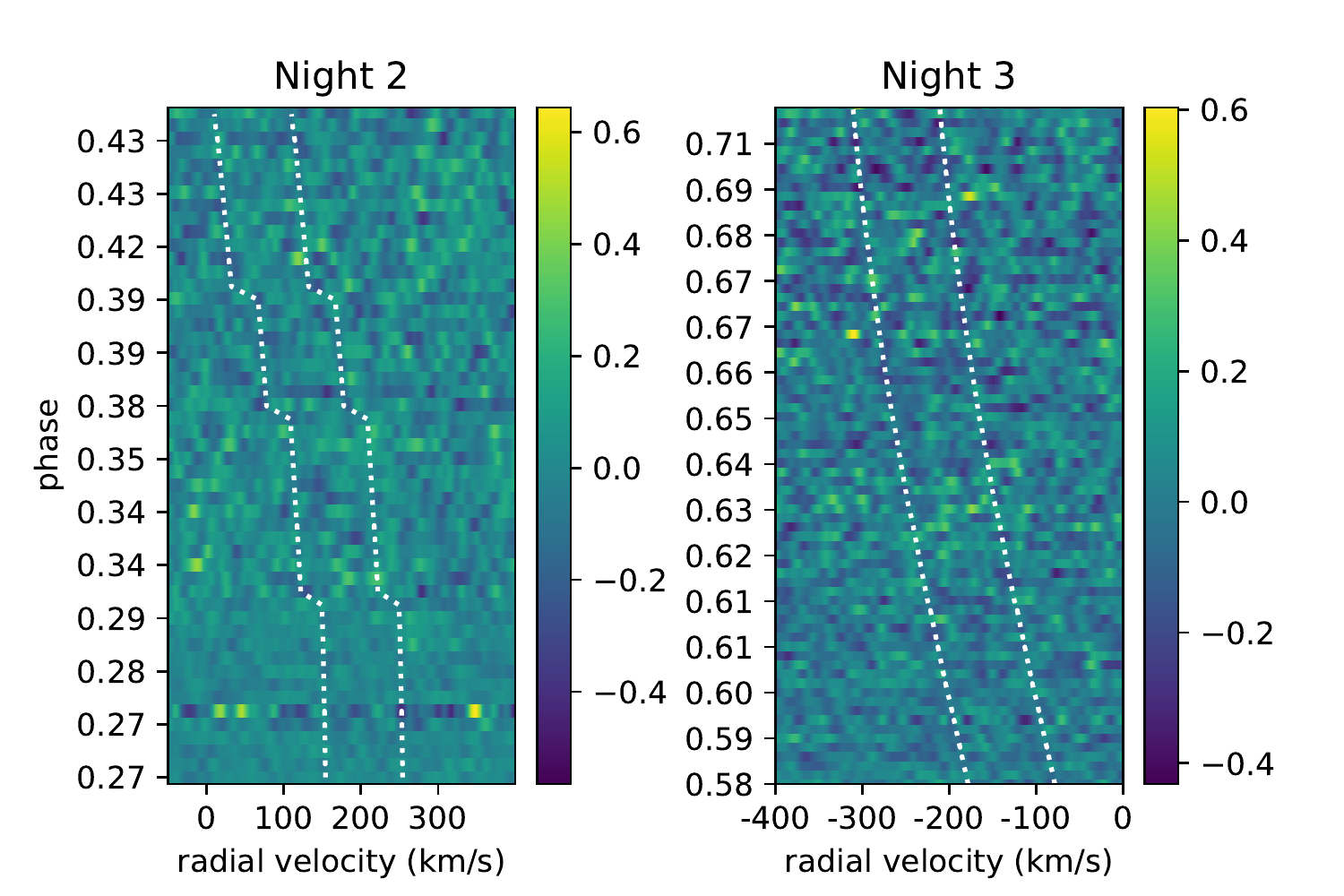} & 
 \includegraphics[width=0.30\textwidth,page=2]{Si_Fossati2021_Solar_for_paper.pdf} & 
  \includegraphics[width=0.30\textwidth,page=3]{Si_Fossati2021_Solar_for_paper.pdf} \\
  
  \end{tabular}
\caption{Cross-correlations from top to bottom of Fe I, Ca II, Mg I, and Si I respectively. Left: the cross-correlation signal as a function of KELT-9b's orbital phase (vertical axis) and systemic velocity (horizontal axis) for Nights 2 and 3. The planet signal is in the middle of the white dotted guiding lines.  Middle: The phase-folded cross-correlation signal as a function of planet orbital velocity ($K_p$) and systemic velocity. The vertical and horizontal white dotted lines indicate the best systemic velocity and orbital velocity, respectively. Right: the phase-folded cross-correlation signal at the best value of $K_p$.}
\label{fig:ClassicalCCF}
\end{figure*}

\begin{figure*}[!htb]
\centering
\includegraphics[width=0.85\textwidth]{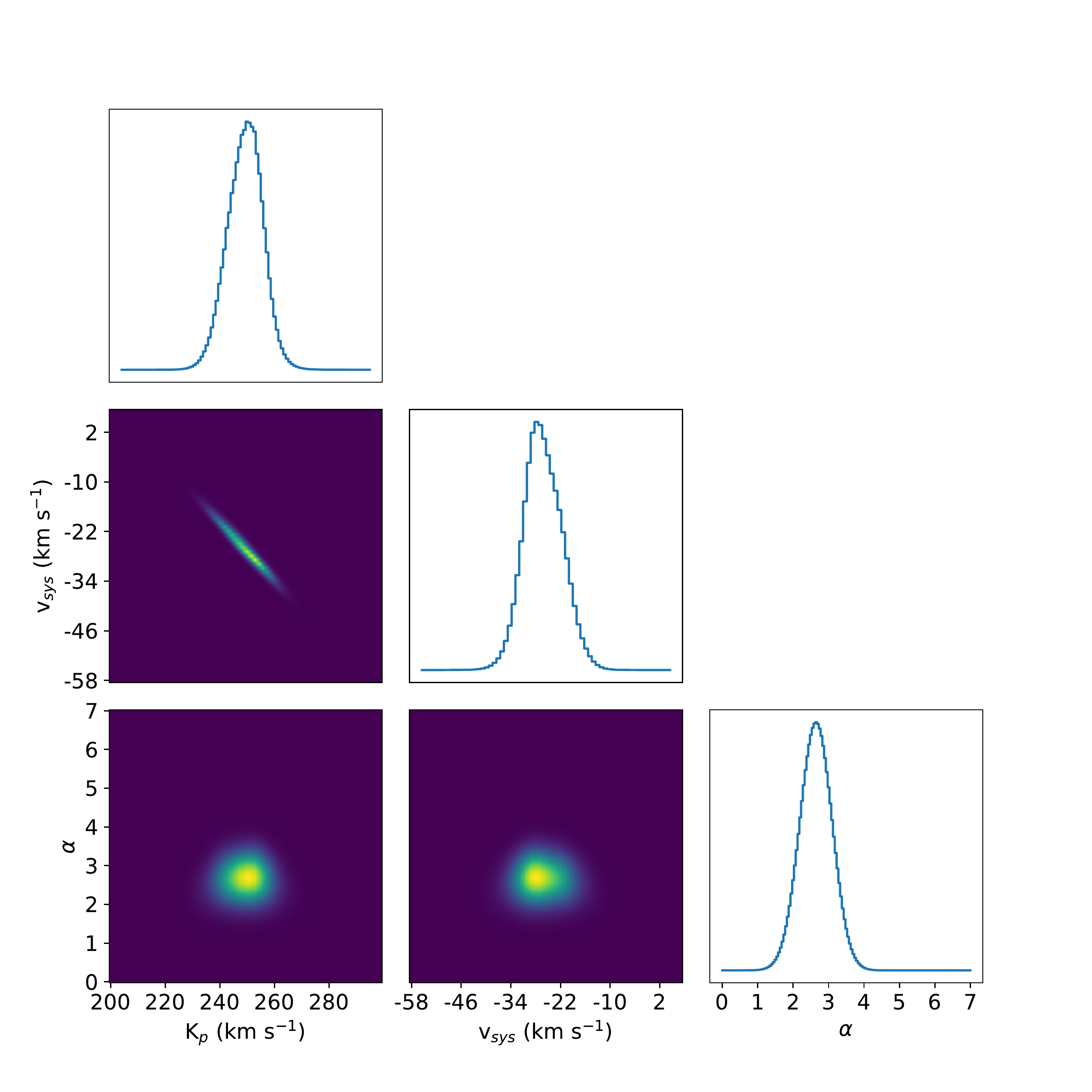} 
\caption{The conditional (2D) and marginalized (1D) likelihood distributions for Fe I using the \citet{Lothringer2018} pressure-temperature profile for Night 2.}
\label{fig:Fe_Lothringer2018_DetSig_lnL_Night2}
\end{figure*}

\begin{figure*}[!htb]
\centering
\includegraphics[width=0.85\textwidth]{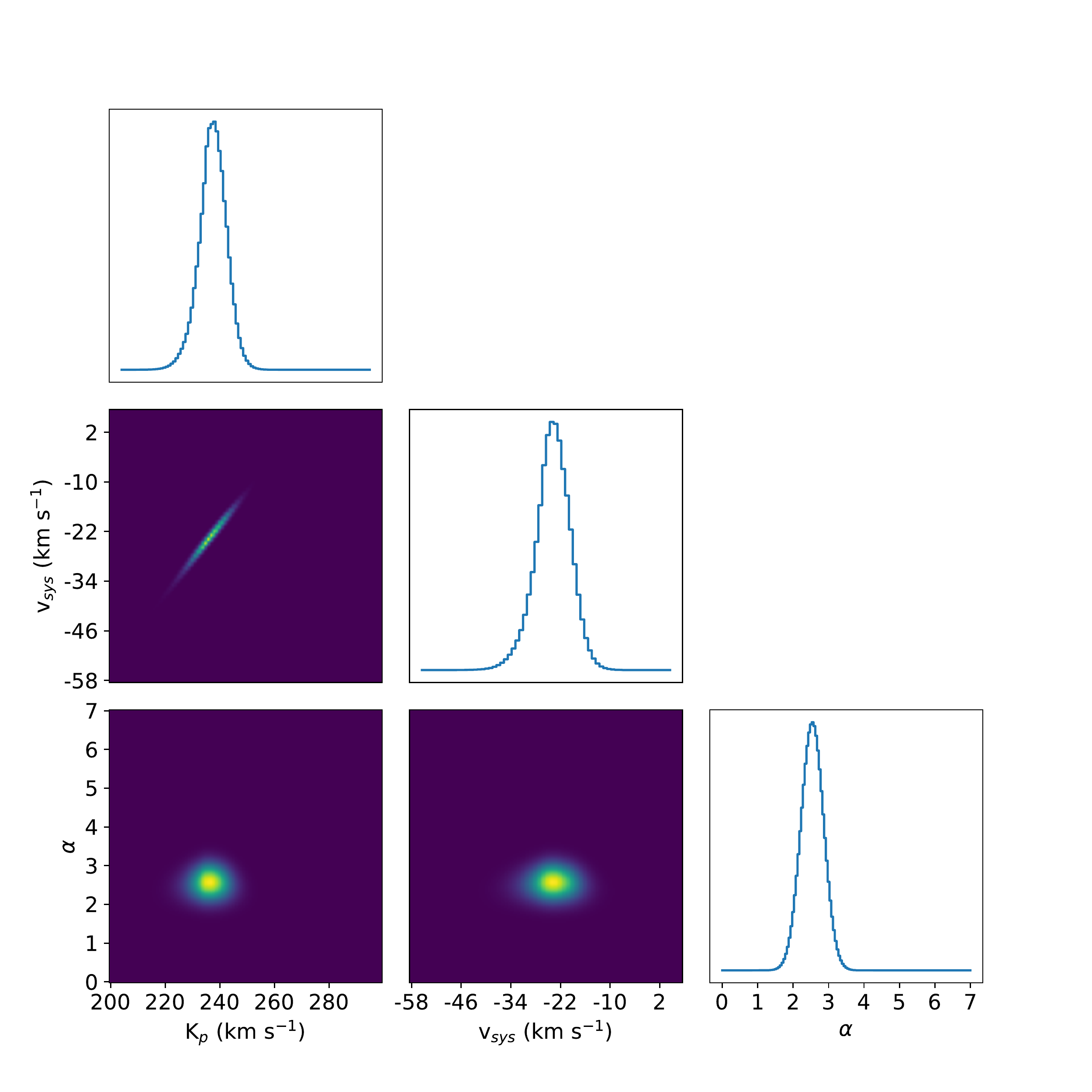} 
\caption{The conditional (2D) and marginalized (1D) likelihood distributions for Fe I using the \citet{Lothringer2018} pressure-temperature profile for Night 3.}
\label{fig:Fe_Lothringer2018_DetSig_lnL_Night3}
\end{figure*}

\begin{figure*}[!htb]
\centering
\includegraphics[width=0.85\textwidth]{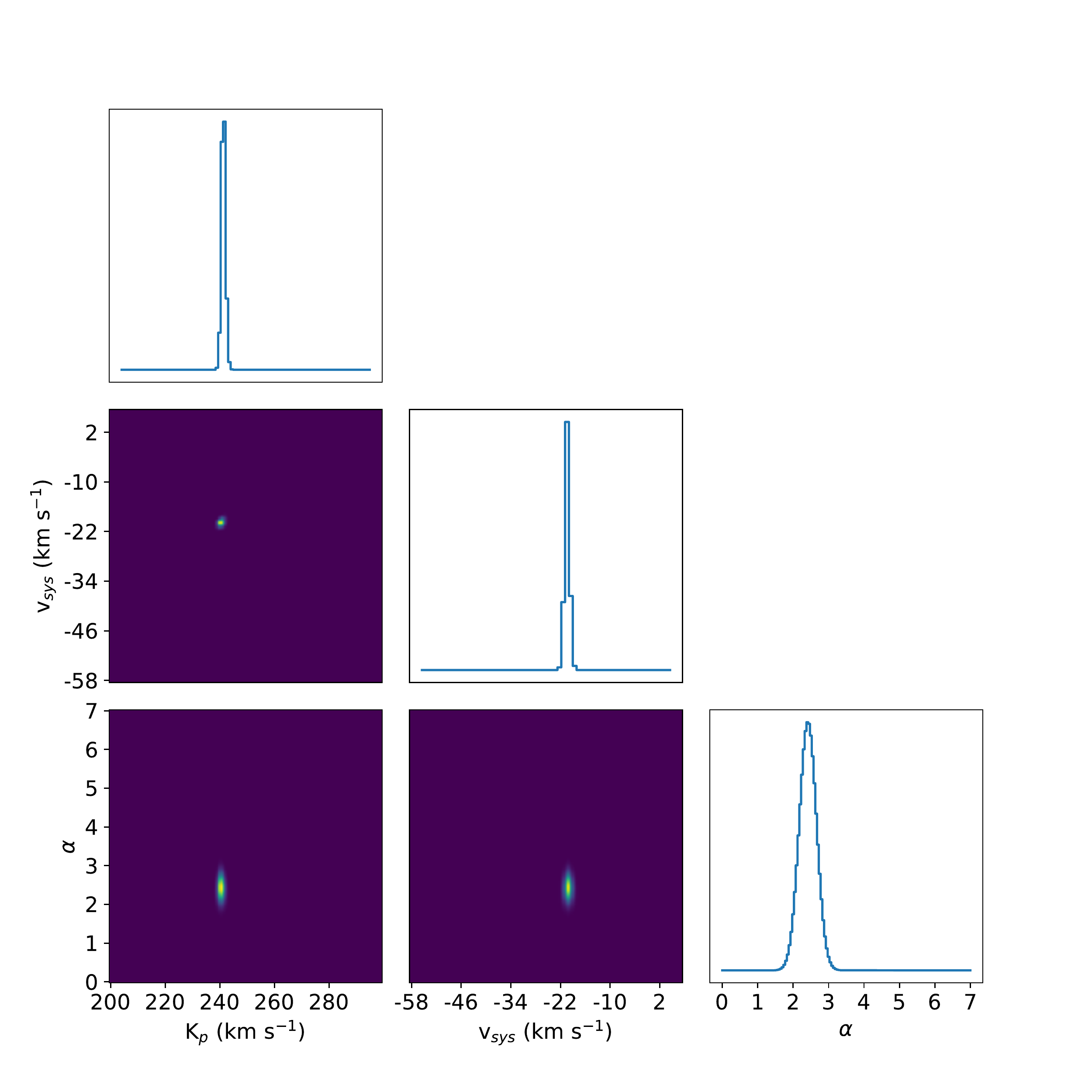} 
\caption{The conditional (2D) and marginalized (1D) likelihood distributions for Fe I using the \citet{Lothringer2018} pressure-temperature profile after combining the log-likelihoods from all nights.}
\label{fig:Fe_Lothringer2018_DetSig_lnL}
\end{figure*}

We found evidence for emission by Fe I, Ca II, Mg I, and Si I. The classical cross-correlation plots of these detections are shown in Fig. \ref{fig:ClassicalCCF} while the constraints on $K_p$, $v_{sys}$, and $\alpha$ derived from the likelihood mapping are shown in Table \ref{tab:SummaryOfDetections}.  For the likelihood fits involving only $K_p$, $v_{sys}$, and $\alpha$, the detection significance is well approximated by the significance to which $\alpha$ is constrained.  An example of the 2D conditional distributions and 1D marginalized distributions to constrain the detection of Fe I using the \citet{Lothringer2018} $P$-$T$ profile are shown in Figs. \ref{fig:Fe_Lothringer2018_DetSig_lnL_Night2}, \ref{fig:Fe_Lothringer2018_DetSig_lnL_Night3}, and \ref{fig:Fe_Lothringer2018_DetSig_lnL} for Nights 2, 3 and all nights combined, respectively.

%%%%%%%%%%%%%%%%%%%

%%%%%%%%%%%%%%%%%%%%

%%%%%%%%%%%%%%%%%%%

We find evidence for Si I emission to 4.01$\sigma$ and 4.10 $\sigma$ using the \citet{Lothringer2018} and \citet{Fossati2021} $P$-$T$ profiles, respectively. This is the first evidence for Si I in the atmosphere of KELT-9b and the third finding of evidence of Si I emission from any exoplanet \citep[the first two being WASP-33b and KELT-20b/MASCARA-2b;][]{Cont2022}.  Theoretical modeling of ultrahot Jupiters suggests that atomic Si I may be present on their hot daysides, while Si-bearing molecules such as SiO may be present on their cooler nightsides and terminators \citep[e.g.,][]{Helling2019}. Such a scenario may reconcile our potential detection of Si I in emission with the non-detection in transmission by \citet{Hoeijmakers2019}. As SiO was recently detected in the transmission spectrum of the cooler ultrahot Jupiter WASP-178 b \citep{Lothringer2022}, it should also be searched for in the transmission spectrum of KELT-9b as its detection in transmission, along with the Si I detection in emission, could allow Si chemistry to be studied as a function of longitude.  We did not search for Si II as its opacity data are not included in \texttt{petitRADTRANS} by default.  However, its detection may also provide insight into KELT-9b's Si chemistry so it should also be searched for in future work.

Unlike Si I, Ca II and Mg I have been previously detected in transmission by \citet{Turner2020} and \citet{Hoeijmakers2019}, respectively. Therefore, our emission detections allow KELT-9b's dayside and terminator to be compared. As a cross-correlation function approximates an average line profile, the width of the signal can indicate broadening processes. To measure the broadening, we fit Gaussians to the phase-folded combined CCFs at the best-fit values of $K_p$ and $v_{sys}$ from the likelihood mapping. These widths, expressed in FHWM, are shown in Table \ref{tab:CCFBroadening}, along with FHWMs reported in the literature from transmission spectra.  KELT-9b's tidally locked orbit gives it a $v \sin(i)$ of approximately 6 km s$^{-1}$.  Therefore, the CCFs indicating robust detections are expected to have FWHMs of at least 2$\times v \sin(i) = 12$ km s$^{-1}$.  For Mg I, the FWHM in emission and transmission are consistent to within 1$\sigma$, indicating similar broadening mechanisms. For Ca II, the FHWM is much wider in emission than transmission, possibly indicating increased thermal Doppler broadening from the hot dayside \citep[e.g.,][]{Hill2013} or increased pressure broadening from the higher pressures probed in emission \citep[e.g.][]{Hedges2016}. However, as the evidence for Ca II emission is only at the 3.1$\sigma$ level, this interpretation should be confirmed with additional observations.

%%%%%%%%%%%%%%%%%%
\begin{table*}[!htb]
\caption{\label{tab:SummaryOfDetections} Summary of detections}
\begin{tabular}{llllll}
\hline \hline
Model & Parameter    & All nights                                  & Nights 2 and 3                              & Night 2                                      & Night 3                                     \\ \hline 
G: Fe I & $\alpha$     & 0.84 $_{-0.08}^{+0.09}$ (9.66 $\sigma$)     & 0.91 $_{-0.09}^{+0.09}$ (9.87 $\sigma$)     & 0.96 $_{-0.16}^{+0.16}$ (6.10 $\sigma$)      & 0.88 $_{-0.11}^{+0.11}$ (7.80 $\sigma$)     \\
      & $K_p$        & 240.75 $_{-0.75}^{+0.94}$                   & 240.50 $_{-0.92}^{+0.95}$                   & 245.12 $_{-6.18}^{+6.82}$                    & 239.83 $_{-5.35}^{+5.45}$                   \\
      & $v_{sys}$    & -19.59 $_{-0.83}^{+0.61}$                   & -20.10 $_{-0.68}^{+0.82}$                   & -23.49 $_{-5.30}^{+4.64}$                    & -20.59 $_{-4.51}^{+4.57}$                   \\
      & max $\ln(L)$ & -1.506755e+06                               & -5.788402e+05                               & -3.411358e+05                                & -2.377040e+05                               \\
      & BIC          & 3.013565e+06                                & 1.157733e+06                                & 6.823214e+05                                 & 4.754589e+05                                \\
      & N            & 6.449825e+07                                & 3.971291e+07                                & 1.593439e+07                                 & 2.377852e+07                                \\
L: Fe I & $\alpha$     & 2.39 $_{-0.24}^{+0.24}$ (9.92 $\sigma$)     & 2.55 $_{-0.26}^{+0.26}$ (9.89 $\sigma$)     & 2.62 $_{-0.45}^{+0.45}$ (5.79 $\sigma$)      & 2.52 $_{-0.32}^{+0.32}$ (7.94 $\sigma$)     \\
      & $K_p$        & 240.12 $_{-0.88}^{+0.81}$                   & 240.17 $_{-0.90}^{+0.81}$                   & 249.03 $_{-7.32}^{+6.53}$                    & 236.00 $_{-5.36}^{+5.19}$                   \\
      & $v_{sys}$    & -20.48 $_{-0.62}^{+0.68}$                   & -20.66 $_{-0.84}^{+0.59}$                   & -27.01 $_{-4.56}^{+5.36}$                    & -24.10 $_{-4.47}^{+4.28}$                   \\
      & max $\ln(L)$ & -1.506758e+06                               & -5.788443e+05                               & -3.411382e+05                                & -2.377047e+05                               \\
      & BIC          & 3.013569e+06                                & 1.157741e+06                                & 6.823261e+05                                 & 4.754604e+05                                \\
F: Fe I & $\alpha$     & 3.23 $_{-0.33}^{+0.33}$ (9.73 $\sigma$)     & 3.41 $_{-0.36}^{+0.36}$ (9.57 $\sigma$)     & 3.42 $_{-0.63}^{+0.63}$ (5.42 $\sigma$)      & 3.41 $_{-0.44}^{+0.44}$ (7.79 $\sigma$)     \\
      & $K_p$        & 240.10 $_{-0.88}^{+0.84}$                   & 240.12 $_{-0.91}^{+0.87}$                   & 249.54 $_{-8.12}^{+7.22}$                    & 234.78 $_{-5.68}^{+5.28}$                   \\
      & $v_{sys}$    & -20.59 $_{-0.80}^{+0.57}$                   & -20.77 $_{-0.87}^{+0.66}$                   & -27.35 $_{-5.00}^{+5.85}$                    & -25.14 $_{-4.77}^{+4.33}$                   \\
      & max $\ln(L)$ & -1.506760e+06                               & -5.788477e+05                               & -3.411401e+05                                & -2.377059e+05                               \\
      & BIC          & 3.013574e+06                                & 1.157748e+06                                & 6.823300e+05                                 & 4.754627e+05                                \\ \hline
L: Ca II & $\alpha$     & 0.62 $_{-0.20}^{+0.18}$ (3.13 $\sigma$)   & 0.67 $_{-0.21}^{+0.19}$ (3.20 $\sigma$)     & 0.86 $_{-0.41}^{+0.37}$ (2.10 $\sigma$)      & 0.56 $_{-0.28}^{+0.24}$ (2.02 $\sigma$)     \\
      & $K_p$        & 238.96 $_{-6.82}^{+11.19}$                  & 242.41 $_{-7.70}^{+8.32}$                   & 270.21 $_{-59.00}^{+38.60}$                  & 278.41 $_{-51.99}^{+46.92}$                 \\
      & $v_{sys}$    & -13.94 $_{-4.89}^{+9.57}$                   & -11.28 $_{-6.00}^{+6.47}$                   & -38.46 $_{-23.93}^{+42.43}$                  & 17.47 $_{-30.39}^{+34.25}$                  \\
      & max $\ln(L)$ & -1.506799e+06                               & -5.788854e+05                               & -3.411502e+05                                & -2.377322e+05                               \\
      & BIC          & 3.013651e+06                                & 1.157823e+06                                & 6.823501e+05                                 & 4.755153e+05                                \\
F: Ca II & $\alpha$     & 0.57 $_{-0.17}^{+0.16}$ (3.30 $\sigma$)   & 0.62 $_{-0.18}^{+0.17}$ (3.44 $\sigma$)     & 0.79 $_{-0.38}^{+0.33}$ (2.10 $\sigma$)      & 0.56 $_{-0.24}^{+0.21}$ (2.37 $\sigma$)     \\
      & $K_p$        & 238.31 $_{-5.57}^{+11.65}$                  & 242.26 $_{-6.96}^{+7.50}$                   & 272.15 $_{-56.24}^{+32.25}$                  & 275.68 $_{-35.79}^{+46.19}$                 \\
      & $v_{sys}$    & -14.43 $_{-4.26}^{+9.81}$                   & -11.18 $_{-5.50}^{+5.91}$                   & -39.97 $_{-19.22}^{+41.13}$                  & 15.17 $_{-23.75}^{+30.21}$                  \\
      & max $\ln(L)$ & -1.506798e+06                               & -5.788848e+05                               & -3.411497e+05                                & -2.377311e+05                               \\
      & BIC          & 3.013650e+06                                & 1.157822e+06                                & 6.823492e+05                                 & 4.755131e+05                                \\ \hline
L: Mg I & $\alpha$     & 3.73 $_{-0.67}^{+0.66}$ (5.55 $\sigma$)     & 4.17 $_{-0.71}^{+0.70}$ (5.89 $\sigma$)     & 5.10 $_{-1.59}^{+1.50}$ (3.20 $\sigma$)      & 4.35 $_{-0.81}^{+0.80}$ (5.38 $\sigma$)     \\
      & $K_p$        & 242.81 $_{-2.35}^{+2.39}$                   & 243.68 $_{-3.47}^{+2.53}$                   & 277.04 $_{-14.59}^{+25.79}$                  & 227.99 $_{-9.09}^{+9.82}$                   \\
      & $v_{sys}$    & -17.34 $_{-2.06}^{+2.07}$                   & -17.24 $_{-2.95}^{+2.15}$                   & -39.00 $_{-12.07}^{+10.34}$                  & -30.33 $_{-7.22}^{+8.22}$                   \\
      & max $\ln(L)$ & -1.506790e+06                               & -5.788743e+05                               & -3.411480e+05                                & -2.377213e+05                               \\
      & BIC          & 3.013635e+06                                & 1.157801e+06                                & 6.823457e+05                                 & 4.754935e+05                                \\
F: Mg I & $\alpha$     & 5.92 $_{-1.01}^{+1.00}$ (5.88 $\sigma$)     & 6.59 $_{-1.06}^{+1.06}$ (6.19 $\sigma$)     & 8.13 $_{-2.34}^{+2.22}$ (3.48 $\sigma$)      & 6.71 $_{-1.22}^{+1.21}$ (5.52 $\sigma$)     \\
      & $K_p$        & 243.54 $_{-2.32}^{+2.27}$                   & 244.65 $_{-2.98}^{+2.11}$                   & 277.45 $_{-14.10}^{+26.17}$                  & 230.11 $_{-8.75}^{+9.63}$                   \\
      & $v_{sys}$    & -16.92 $_{-2.01}^{+1.89}$                   & -16.55 $_{-2.49}^{+1.84}$                   & -39.46 $_{-12.15}^{+10.29}$                  & -28.75 $_{-7.09}^{+8.13}$                   \\
      & max $\ln(L)$ & -1.506789e+06                               & -5.788727e+05                               & -3.411472e+05                                & -2.377208e+05                               \\
      & BIC          & 3.013632e+06                                & 1.157798e+06                                & 6.823442e+05                                 & 4.754926e+05                                \\ \hline
L: Si I & $\alpha$     & 2.44 $_{-0.61}^{+0.59}$ (4.01 $\sigma$)     & 1.86 $_{-0.83}^{+0.68}$ (2.24 $\sigma$)     & 2.18 $_{-1.45}^{+1.31}$ (1.50 $\sigma$)      & 2.03 $_{-0.97}^{+0.82}$ (2.09 $\sigma$)     \\
      & $K_p$        & 241.94 $_{-2.06}^{+23.35}$                  & 267.05 $_{-28.52}^{+56.20}$                 & 311.20 $_{-48.27}^{+18.06}$                  & 235.49 $_{-20.87}^{+54.00}$                 \\
      & $v_{sys}$    & -18.45 $_{-1.74}^{+4.00}$                   & -17.89 $_{-3.40}^{+22.63}$                  & -15.77 $_{-22.87}^{+13.41}$                  & -25.31 $_{-15.67}^{+22.73}$                 \\
      & max $\ln(L)$ & -1.506797e+06                               & -5.788872e+05                               & -3.411514e+05                                & -2.377319e+05                               \\
      & BIC          & 3.013648e+06                                & 1.157827e+06                                & 6.823526e+05                                 & 4.755148e+05                                \\
F: Si I & $\alpha$     & 3.87 $_{-0.94}^{+0.91}$ (4.10 $\sigma$)     & 2.98 $_{-1.27}^{+1.06}$ (2.35 $\sigma$)     & 3.54 $_{-2.26}^{+1.99}$ (1.57 $\sigma$)      & 3.18 $_{-1.49}^{+1.28}$ (2.13 $\sigma$)     \\
      & $K_p$        & 242.06 $_{-2.11}^{+31.15}$                  & 266.92 $_{-28.43}^{+56.15}$                 & 311.18 $_{-47.06}^{+17.31}$                  & 236.52 $_{-21.86}^{+56.15}$                 \\
      & $v_{sys}$    & -18.36 $_{-1.78}^{+24.63}$                  & -17.87 $_{-3.28}^{+22.96}$                  & -15.50 $_{-22.37}^{+12.90}$                  & -24.80 $_{-16.36}^{+22.81}$                 \\
      & max $\ln(L)$ & -1.506797e+06                               & -5.788871e+05                               & -3.411513e+05                                & -2.377320e+05                               \\
      & BIC          & 3.013648e+06                                & 1.157827e+06                                & 6.823524e+05                                 & 4.755150e+05                                 \\
      & N            & 6.449825e+07                                & 3.971291e+07                                & 1.593439e+07                                 & 2.377852e+07                                \\ \hline
\multicolumn{6}{l}{Note 1: The number of data points, $N$, only depends on the night.} \\
\multicolumn{6}{l}{Note 2: These BIC calculations all use $k=3$ for the three model parameters $\alpha$, $K_p$, and $v_{sys}$.} \\
\multicolumn{6}{l}{Note 3: The values in parentheses are the significance to which $\alpha$ is constrained. These values} \\
\multicolumn{6}{l}{~~~~~~~~~~~ also indicate the detection significance.}
\end{tabular}
\end{table*}

\clearpage

%%%%%%%%%%%%%%%%%%%%%

We robustly detected neutral iron in emission, confirming the previous detections of Fe I emission at high-resolution by \cite{Pino2020,Pino2022} and \cite{Kasper2021}.

 \begin{table*}[!htb]
 \centering
\caption{\label{tab:CCFBroadening} Detected Signal Broadening}
\begin{tabular}{ccccc} \hline  \hline

 &  \citet{Lothringer2018} &   \citet{Fossati2021}  &   \\ 
 &  $P$-$T$ Profile &   $P$-$T$ Profile  & \multicolumn{2}{c}{Transmission}  \\ 
 & FWHM  & FWHM  & FWHM &  \\ 
Species & (km s$^{-1}$) & (km s$^{-1}$) & (km s$^{-1}$) & Source \\ \hline
Fe I    & 14.72 $\pm$ 0.64 & 15.05 $\pm$ 0.66 & 20.1 $\pm$ 1.0 & \citep{Hoeijmakers2019} \\
Ca II & 61.69 $\pm$ 3.70 & 53.80 $\pm$ 3.34 & 22.1$^{+1.6}_{-1.6}$ $^\dagger$ & \citep{Turner2020} \\
Mg I    & 22.90 $\pm$ 1.98 & 24.21 $\pm$ 2.03 & 27.5 $\pm$ 4.3 & \citep{Hoeijmakers2019} \\
Si I    & 12.82 $\pm$ 1.22 & 12.97 $\pm$ 1.23 & $-$ & $-$ \\
\hline
\multicolumn{5}{c}{$\dagger$ An average of the FWHM fits to the three Ca II lines, presented by \citet{Turner2020}.}
\end{tabular}
\end{table*}

 %%%%%%%%%%

The variation in velocities across the detected species is similar to that reported by \cite{Hoeijmakers2019} from transmission spectra.  Our median systemic velocity is $-$17.55$^{+2.51}_{-1.27}$ which is close to the systemic velocity reported by \cite{Hoeijmakers2019} of $-$17.74 $\pm$ 0.11 km s$^{-1}$ (see Table \ref{tab:LiteratureKpVsys}).  However, if we only consider neutral iron with the \citet{Lothringer2018} $P$-$T$ profile, which is our strongest detection, we find a systemic velocity of $-$20.48$^{+0.68}_{-0.62}$ which is more consistent with the values reported by \cite{Gaudi2017}, \cite{Borsa2019}, and \cite{Pino2020} (see Table \ref{tab:LiteratureKpVsys}).

\begin{table*}[!htb]
\centering
\caption{\label{tab:LiteratureKpVsys} Literature and Derived $v_{sys}$ and $K_p$ Values}
\begin{tabular}{ccc} \hline  \hline
 & $v_{sys}$ (km s$^{-1}$) & $K_P$ \\ \hline
\cite{Gaudi2017} & $-$20.6 $\pm$ 0.1 & 269$^{+6.5}_{-6}$ \\
\cite{Borsa2019} & $-$19.81 $\pm$ 0.02 & ... \\
\cite{Hoeijmakers2019} & $-$17.7 $\pm$ 0.1 & 234.24 $\pm$ 0.9 \\

\cite{Pino2020} & $-$20.5$^{+2}_{-1.5}$ & 241.5$^{+3}_{-2}$ \\

This work & $-$20.48$^{+0.68}_{-0.62}$ & 240.12$^{+0.81}_{-0.88}$ \\

\hline
\end{tabular}
\end{table*}

\subsection{Nondetected Species} 

We found no evidence of dayside emission from Al I, Ca I, Cr I, Fe II, FeH, K I, Li I, Mg II, Na I, OH, Ti I, TiO, V I, V II, VO, \& Y I.   \citet{Pino2020} also did not detect Fe II emission but did detect neutral iron emission and constrain it to form at a pressure of 10$^{-3}$ $-$ 10$^{-5}$ bar. From transmission spectroscopy, \citet{Hoeijmakers2019} estimate that their detected Fe II absorption occurs at the 10$^{-6}$ bar level. Therefore, \citet{Pino2020} suggest that their non-detection of Fe II emission results from Fe II primarily forming at low pressures that are challenging to detect with emission spectroscopy.  This may also be the case for other ionized species such as Mg II as the models of \citet{Kitzmann2018} generally indicate that ionized species have lower abundances at higher pressures.

To assess the sensitivity of our data to the species we did not detect, we injected model spectra into the data and tried to recover the injected signals. We were unable to recover any injected signals, indicating that data with a higher signal-to-noise ratio are needed to detect these species.

We found no evidence for emission by the molecular species TiO, in line with \cite{Kasper2021} who derived an upper limit on the TiO VMR that is inconsistent with the VMR reported by \cite{Changeat2021}.

\subsection{Constraining KELT-9b's Atmospheric Properties}

In this section, we use the likelihood mapping to quantitatively constrain KELT-9b's atmosphere.  We first examine whether either the \citet{Lothringer2018} or \cite{Fossati2021} $P$-$T$ profiles are preferred by our data according to the BIC values shown in Table \ref{tab:SummaryOfDetections}. For Fe I emission, we find that $\Delta \mathrm{BIC}_{\mathrm{FeI}} = \mathrm{BIC}_{\mathrm{Foss:FeI}} - \mathrm{BIC}_{\mathrm{Loth:FeI}}$ = 5, indicating moderate evidence for the \citet{Lothringer2018} $P$-$T$ being preferred by our data.  For Mg I we find $\Delta \mathrm{BIC}_{\mathrm{Mg}} =  \mathrm{BIC}_{\mathrm{Loth:Mg}} - \mathrm{BIC}_{\mathrm{Foss:Mg}}$ = 3 indicating weak evidence for the \citet{Fossati2021} $P$-$T$ profile being preferred by our data. For Ca II and Si I we find $\Delta \mathrm{BIC}_{\mathrm{CaII}} = \mathrm{BIC}_{\mathrm{Loth:CaII}} - \mathrm{BIC}_{\mathrm{Foss:CaII}}$ = 1 and $\Delta \mathrm{BIC}_{\mathrm{SiI}} = \mathrm{BIC}_{\mathrm{Loth:SiI}} - \mathrm{BIC}_{\mathrm{Foss:SiI}}$ = 0, respectively, indicating that neither $P$-$T$ profile is preferred by our data. However, the \citet{Fossati2021} $P$-$T$ profile does produce slightly higher detection significances.

\begin{figure*}[!htb]
\centering
\includegraphics[width=0.75\textwidth]{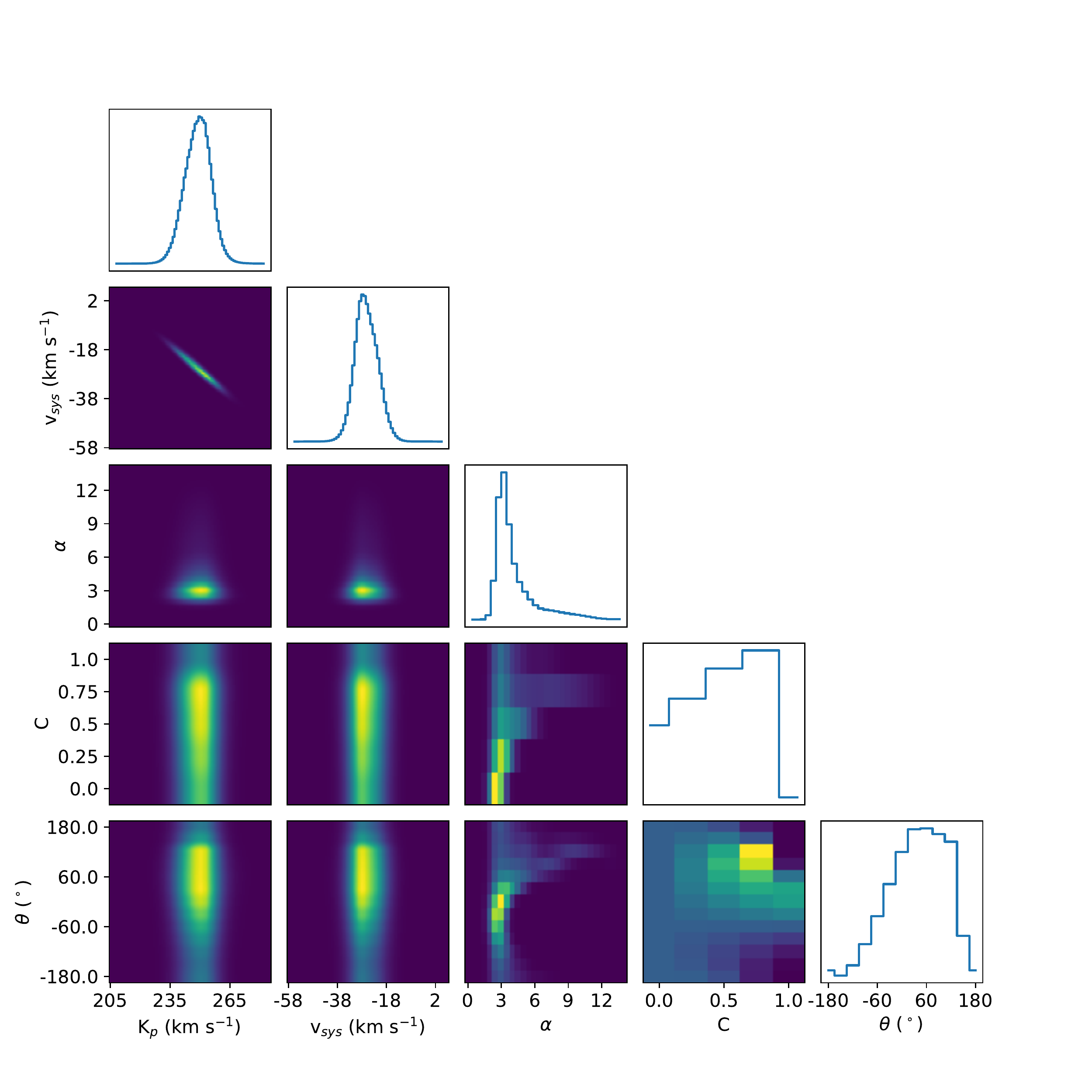} 
\caption{Same as Fig. \ref{fig:Fe_Lothringer2018_DetSig_lnL_Night2} except with the addition of the $\theta$ and $C$ parameters.}
\label{fig:Fe_Lothringer2018_theta_C_lnL_Night2}
\end{figure*}

\begin{figure*}[!htb]
\centering
\includegraphics[width=0.75\textwidth]{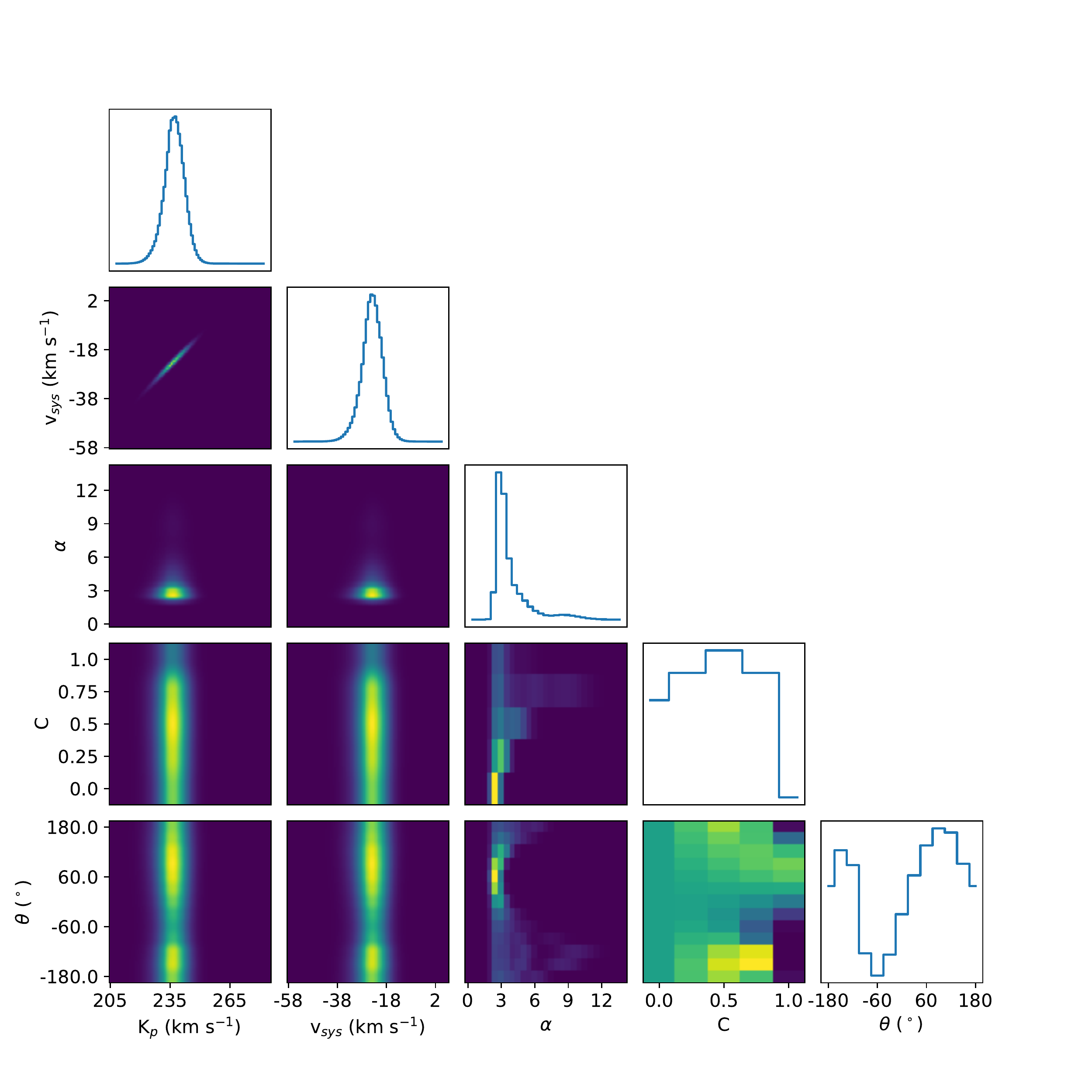} 
\caption{Same as Fig. \ref{fig:Fe_Lothringer2018_DetSig_lnL_Night3} except with the addition of the $\theta$ and $C$ parameters.}
\label{fig:Fe_Lothringer2018_theta_C_lnL_Night3}
\end{figure*}

\begin{figure*}[!htb]
\centering
\includegraphics[width=0.75\textwidth]{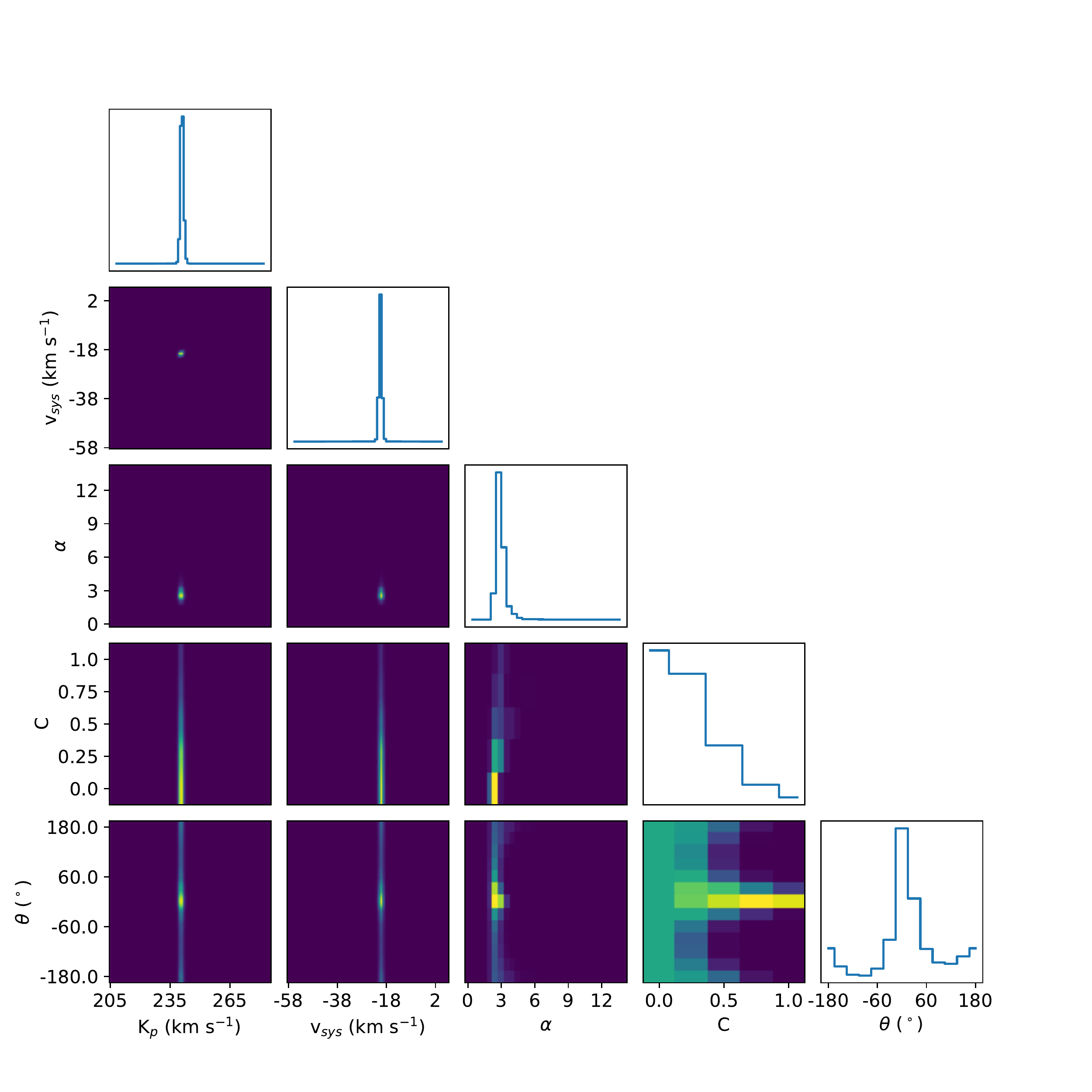} 
\caption{Same as Fig. \ref{fig:Fe_Lothringer2018_DetSig_lnL} except with the addition of the $\theta$ and $C$ parameters.}
\label{fig:Fe_Lothringer2018_theta_C_lnL_all_nights}
\end{figure*}

Next, we add the peak emission offset, $\theta$, and day-night contrast, $C$, parameters to our likelihood analysis for Fe I and Mg I to constrain KELT-9b's 3D emission profile. We focus only on Fe I and Mg I as it was difficult to find robust constraints for the weaker potential detections of Ca II and Si I with these models that include additional terms.  Figures \ref{fig:Fe_Lothringer2018_theta_C_lnL_Night2}, \ref{fig:Fe_Lothringer2018_theta_C_lnL_Night3}, \& \ref{fig:Fe_Lothringer2018_theta_C_lnL_all_nights} show the conditional (2D) and marginalized (1D) likelihood distributions for Fe I using the \citet{Lothringer2018} $P$-$T$ profile, while including $\theta$ and $C$, for Nights 2, 3, and all nights combined, respectively. The derived constraints for both Fe I and Mg I are presented in Table \ref{tab:3DEmissionProfile}.  The addition of $\theta$ and $C$ did not change max $\ln(L)$ so the BIC values increased by 37 and 35 for Fe I and Mg I, respectively.  Therefore, the inclusion of $\theta$ and $C$ is not strictly justified by our data. However, we still present their constraints as they may hint toward additional insights into KELT-9b's atmosphere.  By considering $\theta$, we found that our data could not robustly constrain the phase of peak Fe I and Mg I emission. However, for Fe I there are weak suggestions, particularly in the data from Nights 2 and 3, of the emission peaking slightly after secondary eclipse that may be worth following up with additional observations.  As $C$ is a bounded parameter, we were only able to find upper limits. Table \ref{tab:3DEmissionProfile}, lists the 1-$\sigma$ upper limits of $C$ to be 0.21 and 0.23 for Fe I and Mg I, respectively.  These limits indicate there is little variation in emission line contrast between the day- and nightsides.   For Fe I, we also included all the \citet{Goyal2020} models in our likelihood analysis. However, as these models lack opacity sources that are important for ultrahot Jupiters, any conclusions drawn from these models should be treated with caution.

%%%%%%%%%%%%%%%%%%%%%%%%%%%

%%%%%%%%%%%%%%%%%%%%%%%%%%%%%%%%%%%%%%%%%%%%%

\begin{table*}[!htb]
\centering
\caption{\label{tab:3DEmissionProfile} Constraints on KELT-9b's 3D Emission Profile}
\begin{tabular}{llllll}
\hline \hline
Model & Parameter    & All Nights                                  & Nights 2 and 3                               & Night 2                                     & Night 3                                       \\ \hline
L: Fe I & $\alpha$     & 2.37 $_{-0.31}^{+0.49}$                     & 2.57 $_{-0.41}^{+0.48}$                      & 3.03 $_{-0.79}^{+2.12}$                     & 2.80 $_{-0.62}^{+1.69}$                       \\
      & $K_p$        & 240.11 $_{-0.88}^{+0.83}$                   & 240.16 $_{-0.91}^{+0.82}$                    & 248.94 $_{-7.42}^{+6.65}$                   & 235.97 $_{-5.44}^{+5.25}$                     \\
      & $v_{sys}$    & -20.50 $_{-0.68}^{+0.68}$                   & -20.67 $_{-0.85}^{+0.60}$                    & -26.95 $_{-4.60}^{+5.39}$                   & -24.12 $_{-4.55}^{+4.34}$                     \\
      & $\theta$     & 14.62 $_{-68.41}^{+141.55}$                 & 16.45 $_{-76.76}^{+147.67}$                  & 36.37 $_{-101.96}^{+99.07}$                 & 63.09 $_{-142.78}^{+119.74}$                  \\
      & C $<$ (to 1$\sigma$)       & 0.21                                        & 0.21                                         & 0.41                                        & 0.38                                          \\
      & max $\ln(L)$ & -1.506758e+06                               & -5.788441e+05                                & -3.411377e+05                               & -2.377046e+05                                 \\
      & BIC          & 3.013606e+06                                & 1.157776e+06                                 & 6.823584e+05                                & 4.754941e+05                                  \\ \hline
F: Mg I & $\alpha$     & 3.53 $_{-2.40}^{+3.78}$                     & 6.08 $_{-3.91}^{+2.80}$                      & 11.35 $_{-5.76}^{+14.90}$                   & 8.60 $_{-4.68}^{+9.62}$                       \\ 
      & $K_p$        & 242.23 $_{-3.49}^{+3.02}$                   & 243.35 $_{-4.20}^{+2.05}$                    & 297.46 $_{-30.45}^{+56.43}$                 & 227.57 $_{-9.05}^{+9.83}$                     \\
      & $v_{sys}$    & -17.47 $_{-2.42}^{+2.37}$                   & -17.10 $_{-2.84}^{+1.96}$                    & -49.95 $_{-24.90}^{+15.50}$                 & -30.03 $_{-7.25}^{+7.94}$                     \\
      & $\theta$     & 1.12 $_{-67.64}^{+143.91}$                  & -5.77 $_{-66.55}^{+164.95}$                  & 54.05 $_{-75.81}^{+52.31}$                  & -44.76 $_{-60.18}^{+209.90}$                  \\
      & C $<$ (to 1$\sigma$)     & 0.23                                        & 0.38                                         & 0.65                                        & 0.59                                          \\
      & max $\ln(L)$ & -1.506789e+06                               & -5.788722e+05                                & -3.411459e+05                               & -2.377201e+05                                 \\
      & BIC          & 3.013667e+06                                & 1.157832e+06                                 & 6.823747e+05                                & 4.755250e+05                                  \\
      & N            & 6.449825e+07                                & 3.971291e+07                                 & 1.593439e+07                                & 2.377852e+07  \\ \hline      
\multicolumn{6}{l}{Note: These BIC calculations all use $k=5$ for the five model parameters $\alpha$, $K_p$, $v_{sys}$, $\theta$, and $C$.} \\
\end{tabular}
\end{table*}

%%%%%%%%%%%%%%%%%%%%%%%%%%%%%%%%%%%%%%%%%%%%%%%%%%%%%%%%%

\begin{figure*}[!htb]
\centering
\includegraphics[width=\textwidth]{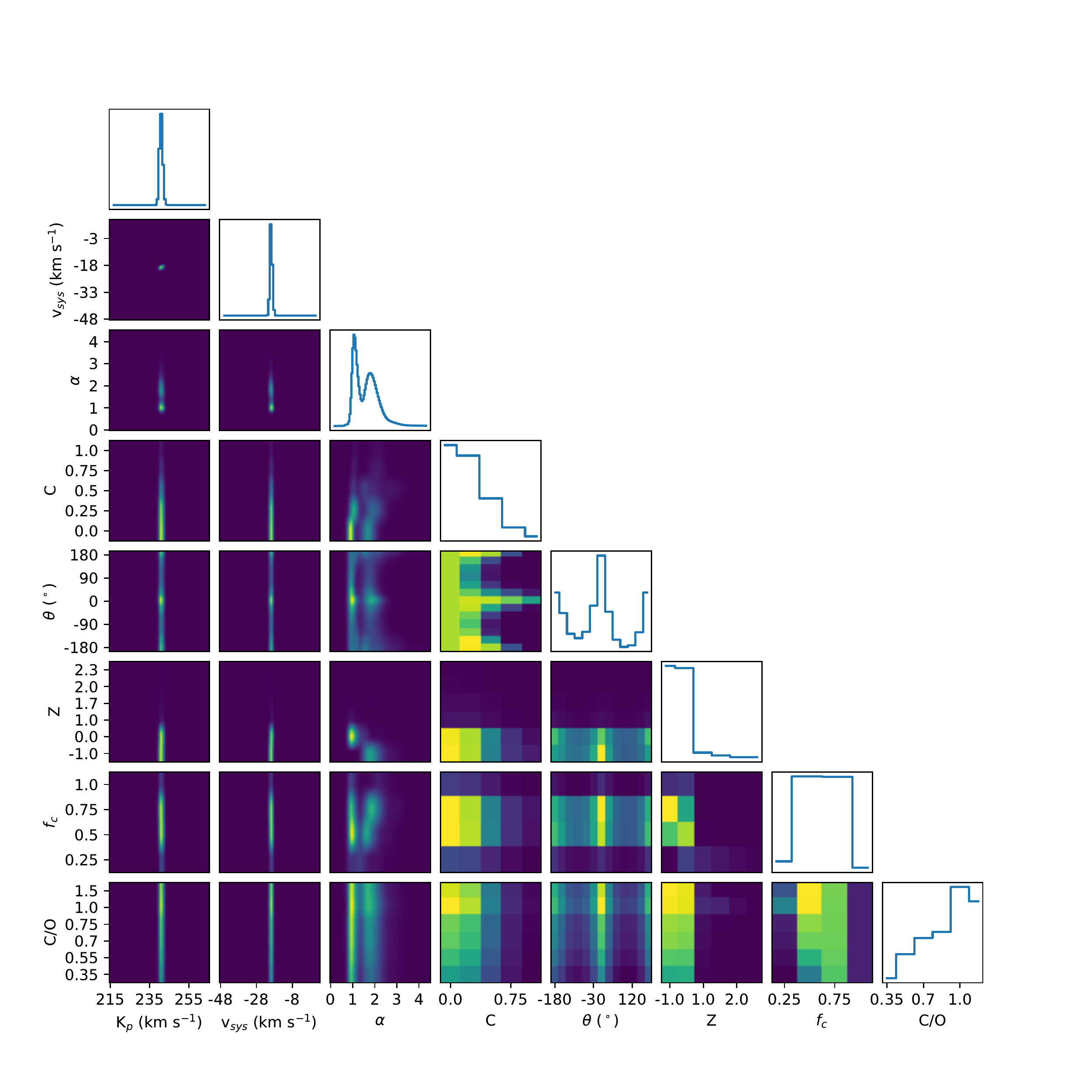} 
\caption{The conditional (2D) and marginalized (1D) likelihood distributions for Fe I using the \citet{Goyal2020} pressure-temperature profile and grid for Night 2.}
\label{fig:CornerPlot618}
\end{figure*}

\begin{figure*}[!htb]
\centering
\includegraphics[width=\textwidth]{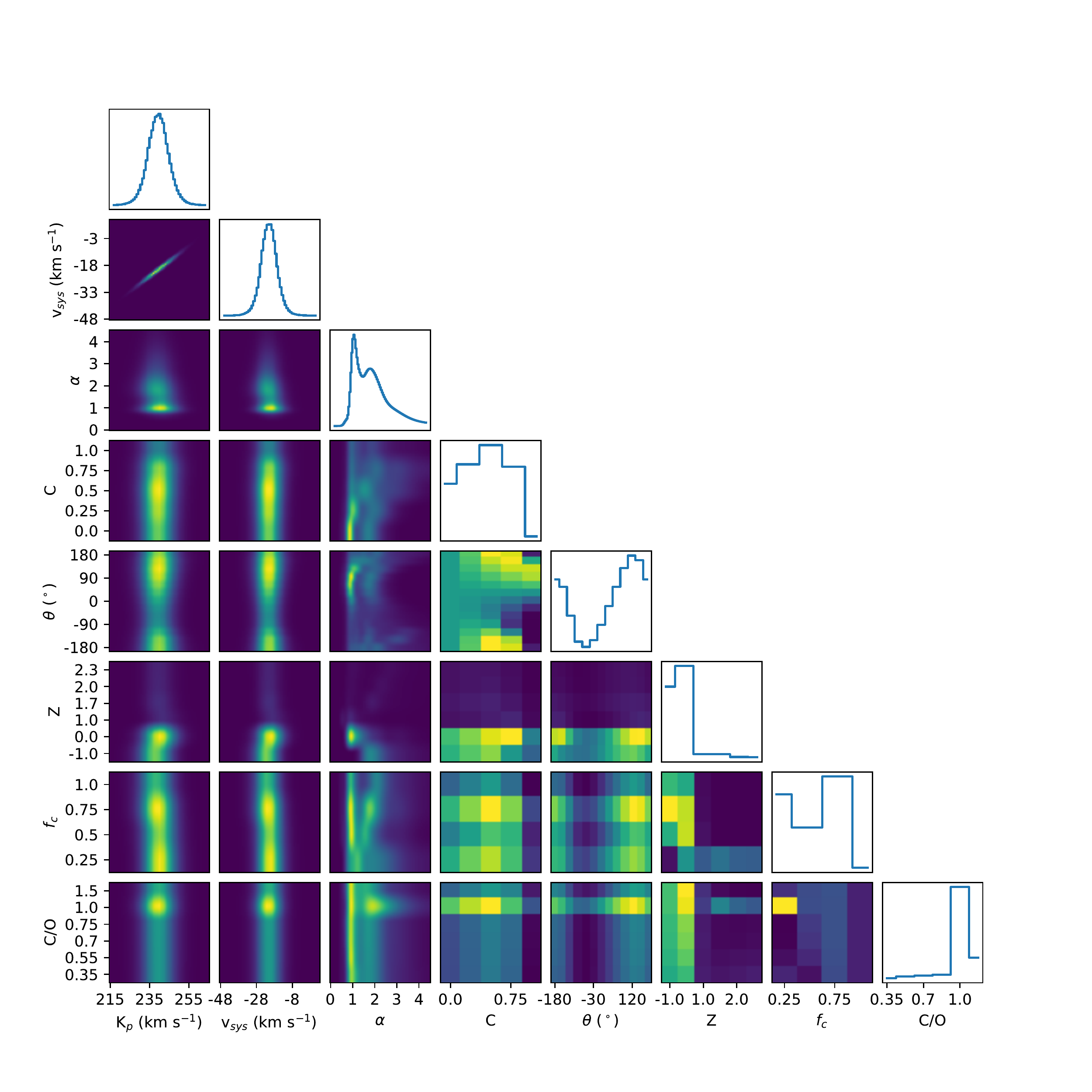} 
\caption{Same as Fig. \ref{fig:CornerPlot618} except for being for Night 3.}
\label{fig:CornerPlot528}
\end{figure*}

\begin{figure*}[!htb]
\centering
\includegraphics[width=\textwidth]{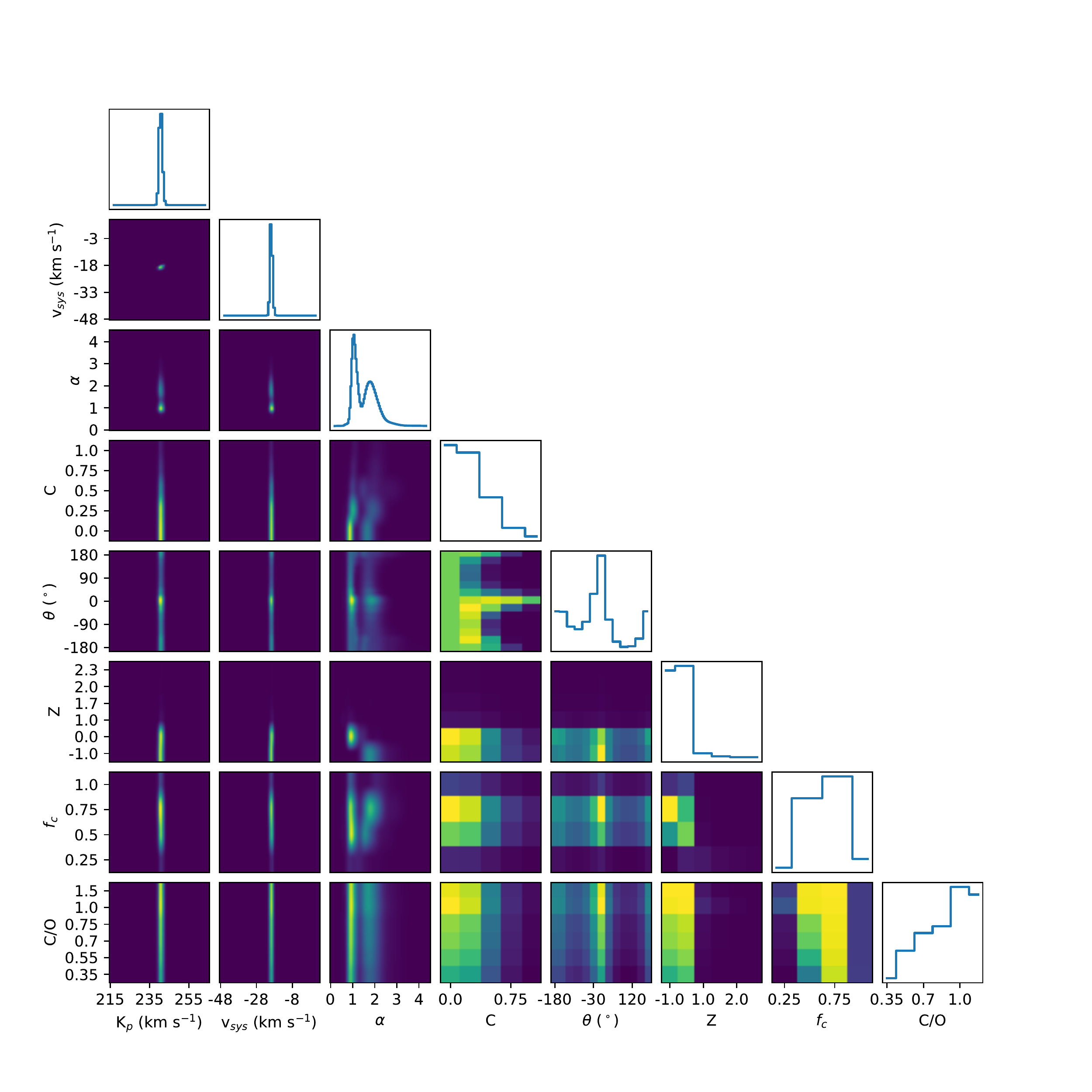} 
\caption{Same as Fig. \ref{fig:CornerPlot618} except for being for all nights combined.}
\label{fig:CornerPlotAllNights}
\end{figure*}

The conditional (2D) and marginalized (1D) likelihood distributions for the \citet{Goyal2020} models for Nights 2, 3, and all nights combined are shown in Figures \ref{fig:CornerPlot618}, \ref{fig:CornerPlot528} \&  \ref{fig:CornerPlotAllNights}, respectively. The best-fit parameters and uncertainties derived from these marginalized distributions are shown in Table \ref{tab:lnL_Fe_Goyal}.  The bimodal distribution of $\alpha$ in these plots may be due to our parameter grid missing informative likelihood structure that occurs between our grid points. Therefore, future research may benefit from utilizing parameterized atmospheric models so that a Markov chain Monte Carlo (MCMC) method can be used to allow more sampling of the parameter space.

\begin{table*}[!htb]
\centering
\caption{\label{tab:lnL_Fe_Goyal} Constraints on KELT-9b's Fe I emission using the \citet{Goyal2020} pressure-temperature profiles.}
\begin{tabular}{llllll}
\hline \hline
Model & Parameter    & All Nights                                  & Nights 2 and 3                               & Night 2                                     & Night 3                                      \\ \hline
G: Fe I & $\alpha$     & 1.40 $_{-0.50}^{+0.60}$                     & 1.21 $_{-0.31}^{+0.70}$                      & 1.50 $_{-0.57}^{+0.56}$                     & 1.65 $_{-0.70}^{+0.97}$                      \\
      & $K_p$        & 240.21 $_{-0.92}^{+0.86}$                   & 240.37 $_{-0.91}^{+0.84}$                    & 240.41 $_{-0.95}^{+0.95}$                   & 238.86 $_{-5.53}^{+5.59}$                    \\
      & $v_{sys}$    & -20.20 $_{-0.64}^{+0.87}$                   & -20.25 $_{-0.59}^{+0.85}$                    & -20.28 $_{-0.61}^{+0.88}$                   & -21.53 $_{-4.66}^{+4.71}$                    \\
      & $\theta$     & 6.63 $_{-86.99}^{+172.74}$                  & 47.94 $_{-119.03}^{+131.72}$                 & 24.27 $_{-97.79}^{+152.95}$                 & 77.39 $_{-141.71}^{+100.97}$                  \\
      & C $<$ (to 1$\sigma$)        & 0.21                                        & 0.19                                         & 0.20                                        & 0.39                           \\
      & Z $<$ (to 1$\sigma$)  & -0.32   & -0.03     & -0.33    & -0.09     \\
      & $f_c$        & 0.54 $_{-0.22}^{+0.18}$                    & 0.48 $_{-0.19}^{+0.20}$                       & 0.49 $_{-0.21}^{+0.21}$                      & 0.49 $_{-0.31}^{+0.31}$                      \\
      & C/O $>$ (to 1$\sigma$)$^\dagger$          & 0.84   & 0.88   & 0.85    & 0.85    \\
      & max $\ln(L)$ & -6.896676e+05                               & -8.165438e+05                                & -5.788400e+05                               & -2.377032e+05                                \\
      & BIC          & 1.379479e+06                                & 1.633228e+06                                 & 1.157813e+06                                & 4.755423e+05                                 \\
      & N            & 6.449825e+07                                & 3.971291e+07                                 & 1.593439e+07                                & 2.377852e+07  \\ \hline
      \multicolumn{6}{l}{Note: These BIC calculations all use $k=8$ for the eight model parameters $\alpha$, $K_p$, $v_{sys}$, $\theta$, $C$, Z, $f_c$, and C/O.} \\
      \multicolumn{6}{l}{$\dagger$ This is not a reliable constraint as Fe I emission depends very weakly on C/O.} \\
\end{tabular}
\end{table*}

%%%%%%%%%%%%%%%%%%%%%%%%%%%%%%%%%

By comparing the BIC values for the \cite{Goyal2020} Fe I models with parameters of $K_p$, $v_{sys}$, and $\alpha$ (see Table \ref{tab:SummaryOfDetections}) and parameters $K_p$, $v_{sys}$, $\alpha$, $\theta$, $C$, Z, $f_c$, and  C/O (see Table \ref{tab:lnL_Fe_Goyal}), we find that the latter models are preferred as they result in a significantly lower BIC ($\Delta \mathrm{BIC} >> 100$).  Using these models, we derived a 1$\sigma$ upper limit on the metallicity to be Z $<$ $-0.32$ ($\approx$ 0.5 $\times$ solar).  While this is a weak constraint, it suggests that KELT-9b's metallicity may be subsolar.  The metallicity of KELT-9b's host star is approximately solar at $-$0.03 $\pm$ 0.20 \citep[][]{Gaudi2017}.   The retrieval of \cite{Changeat2021} constrained KELT-9b's Fe I abundance to be much lower than the solar value. While they consider this to be evidence that Fe exists in other forms in KELT-9b's atmosphere, perhaps this is also partly due to the low intrinsic metallicity suggested by our results. However, their retrieved abundance of TiO suggests a super-solar metallicity. \cite{Kasper2021} constrain the abundance of Fe I to be roughly consistent with solar abundances. However, their constraints on the Fe II abundance imply that KELT-9b's atmospheric metallicity is approximately 100$\times$ solar ([Fe/H] = 2).  As they consider this high metallicity to be unlikely, they instead attribute this to their use of simplifying assumptions in their retrieval.  Similarly, \cite{Pino2020} also constrained KELT-9b's iron abundances to be consistent with its host star (to within a factor of a few).

For the heat recirculation factor, $f_c$, in the \citet{Goyal2020} models, we found that $f_c$ = $0.54_{-0.22}^{+0.18}$ which is consistent with the value reported by \cite{Wong2020} of 0.5\footnote{Reported by \cite{Wong2020} as a recirculation efficiency, $\epsilon$, of 0.39. Converted to a recirculation factor, $f_c$, using $f_c$ = 2/3  - (5/12)$\epsilon$ = 0.504.}. As the C/O ratio should not strongly affect Fe I emission, we do not consider the constraints on C/O suggested by this analysis to be reliable, and include them mostly as an illustration of how unconstrained parameters behave in a log-likelihood analysis. As with the \citet{Lothringer2018} and \citet{Fossati2021} models, the \citet{Goyal2020} models poorly constrain the phase of peak Fe I emission while also indicating little variation in emission line contrast between the day- and nightsides.  Phase curve observations in TESS and Spitzer 4.5 $\mu$m bands indicate that $C$ = 0.8 $-$ 0.9 \citep{Mansfield2020,Wong2020}. Therefore, this suggests that the variation in emission line contrast between the day- and nightsides may differ from the temperature day-night contrast.

The indications from our data of there being little variation in emission line contrast between the day- and nightsides suggests that it may be possible to detect nightside Fe I emission from KELT-9b.  Therefore, KELT-9b may offer the rare opportunity to robustly constrain an exoplanet's nightside $P$-$T$ profile. This would provide observational constraints on how the $P$-$T$ profile changes with planetary longitude, and how far into the nightside it remains inverted. We estimate that it may be possible for Gemini/GRACES to detect KELT-9b's nightside emission. For example, if it were to observe KELT-9b for 6 hr before transit and 6 hr after transit, it could achieve a phase-folded signal-to-noise ratio of over 2500 per pixel.  Therefore, since the cross-correlation method combines the signal from thousands of lines, it would be possible to detect line contrasts well below the continuum nightside emission of 90 ppm found by \citet{Wong2020, Wong2021} from TESS observations.

\section{Conclusion} \label{sec:conclusion}

We searched for atomic and molecular lines in the emission spectrum of KELT-9b by observing partial phase curves with the CARMENES spectrograph. We found evidence for Si I in the atmosphere of KELT-9b for the first time.  Additionally we found evidence for emission by Mg I and Ca II which were previously detected in transmission. We also confirmed previous detections of Fe I emission. Conversely, we find no evidence for dayside emission from Al I, Ca I, Cr I, FeH, Fe II, K I, Li I, Mg II, Na I, OH, Ti I, TiO, V I, V II, VO, Y I.

By employing a cross correlation to likelihood mapping, we found indications of there being little variation in emission line contrast between the day- and nightsides.  This suggests that it may be possible to detect KELT-9b's nightside emission, providing a rare insight into an exoplanet's nightside $P$-$T$ profile.  These results demonstrate that high-resolution ground-based emission spectroscopy can provide valuable insights into exoplanet atmospheres.   

%% IMPORTANT! The old "\acknowledgment" command has be depreciated. It was
%% not robust enough to handle our new dual anonymous review requirements and
%% thus been replaced with the acknowledgment environment. If you try to 
%% compile with \acknowledgment you will get an error print to the screen
%% and in the compiled pdf.
%\begin{acknowledgments}
\section*{Acknowledgments}

We thank Jayesh Goyal for helping us to utilize the atmospheric models presented in \cite{Goyal2020}.  We also thank Paul Mollière for assisting us in our use of \texttt{petitRADTRANS} and Peter Woitke for assisting us in our use of GGchem. We are grateful to Miranda Herman and Stevanus Nugroho for their development of the cross correlation to likelihood mapping code.  We thank Luca Fossati and Joshua Lothringer for providing their KELT-9b $P$-$T$ profiles.   We also thank Jake Turner, Emily Deibert, Ryan MacDonald, and Laura Flagg for helpful discussions.  We thank the anonymous reviewer for their insightful comments that greatly improved our manuscript.

R.K. acknowledges the support by Inter-transfer grant No. LTT-20015. M.K. acknowledges the support from ESA-PRODEX PEA4000127913.

This research has made use of the Extrasolar Planet Encyclopaedia, NASA's Astrophysics Data System Bibliographic Services, and the NASA Exoplanet Archive, which is operated by the California Institute of Technology, under contract with the National Aeronautics and Space Administration under the Exoplanet Exploration Program.

This work is based on observations collected at the Centro Astronómico Hispano Alemán (CAHA), operated jointly by the Max Planck Institut fũr Astronomie and the Instituto de Astrofísica de Andalucia (CSIC)
%\end{acknowledgments}

%% To help institutions obtain information on the effectiveness of their 
%% telescopes the AAS Journals has created a group of keywords for telescope 
%% facilities.
%
%% Following the acknowledgments section, use the following syntax and the
%% \facility{} or \facilities{} macros to list the keywords of facilities used 
%% in the research for the paper.  Each keyword is check against the master 
%% list during copy editing.  Individual instruments can be provided in 
%% parentheses, after the keyword, but they are not verified.

\vspace{5mm}
\facilities{3.5 m Calar Alto telescope/CARMENES.}

%% Similar to \facility{}, there is the optional \software command to allow 
%% authors a place to specify which programs were used during the creation of 
%% the manuscript. Authors should list each code and include either a
%% citation or url to the code inside ()s when available.

\software{\texttt{Astropy} \citep{Astropy2013, Astropy2018}; \texttt{SciPy} \citep{2020SciPy-NMeth}; \texttt{NumPy} \citep{harris2020array}; \texttt{petitRADTRANS} \citep{Molliere2019}; \texttt{Eniric} \citep{Figueira2016}; \texttt{SpectRes} \citep{Carnall2017}.
          }

%% Appendix material should be preceded with a single \appendix command.
%% There should be a \section command for each appendix. Mark appendix
%% subsections with the same markup you use in the main body of the paper.

%% Each Appendix (indicated with \section) will be lettered A, B, C, etc.
%% The equation counter will reset when it encounters the \appendix
%% command and will number appendix equations (A1), (A2), etc. The
%% Figure and Table counter will not reset.

%% Appendix material should be preceded with a single \appendix command.
%% There should be a \section command for each appendix. Mark appendix
%% subsections with the same markup you use in the main body of the paper.

%% Each Appendix (indicated with \section) will be lettered A, B, C, etc.
%% The equation counter will reset when it encounters the \appendix
%% command and will number appendix equations (A1), (A2), etc. The
%% Figure and Table counter will not reset.

%\appendix

%\section{Appendix information}

%% For this sample we use BibTeX plus aasjournals.bst to generate the
%% the bibliography. The sample631.bib file was populated from ADS. To
%% get the citations to show in the compiled file do the following:
%%
%% pdflatex sample631.tex
%% bibtext sample631
%% pdflatex sample631.tex
%% pdflatex sample631.tex

\bibliography{references}{}
\bibliographystyle{aasjournal}

%% This command is needed to show the entire author+affiliation list when
%% the collaboration and author truncation commands are used.  It has to
%% go at the end of the manuscript.
%\allauthors

%% Include this line if you are using the \added, \replaced, \deleted
%% commands to see a summary list of all changes at the end of the article.
%\listofchanges

\end{document}